\documentclass{emulateapj}
\usepackage{amsmath}
\usepackage{graphicx}
\shorttitle{Continuously Driven Ion Velocity Space Instabilities}
\shortauthors{}

\begin{document}
\title{PIC Simulations of Continuously Driven Mirror and Ion Cyclotron Instabilities in High Beta Astrophysical and Heliospheric Plasmas}
\author{Mario A. Riquelme\altaffilmark{1}, Eliot Quataert\altaffilmark{2} \& Daniel Verscharen\altaffilmark{3}}
\altaffiltext{1}{Departamento de F\'isica, Facultad de Ciencias F\'isicas y Matem\'aticas, Universidad de Chile; mario.riquelme@dfi.uchile.cl}
\altaffiltext{2}{Astronomy Department and Theoretical Astrophysics Center, University of California, Berkeley, CA 94720; eliot@berkeley.edu}
\altaffiltext{3}{Space Science Center and Department of Physics, University of New Hampshire, Durham, NH 03824; Daniel.Verscharen@unh.edu}

\begin{abstract} \noindent We use particle-in-cell (PIC) simulations to study the nonlinear evolution of ion velocity space instabilities in an idealized problem in which a background velocity shear continuously amplifies the magnetic field. We simulate the astrophysically relevant regime where the shear timescale is long compared to the ion cyclotron period, and the plasma beta is $\beta \sim 1-100$. The background field amplification in our calculation is meant to mimic processes such as turbulent fluctuations or MHD-scale instabilities.  The field amplification continuously drives a pressure anisotropy with $p_\perp > p_\parallel$ and the plasma becomes unstable to the mirror and ion cyclotron instabilities.  In all cases, the nonlinear state is dominated by the mirror instability, not the ion cyclotron instability, and the plasma pressure anisotropy saturates near the threshold for the linear mirror instability. The magnetic field fluctuations initially undergo exponential growth but saturate in a secular phase  in which the fluctuations grow on the same timescale as the background magnetic field (with $\delta B \sim 0.3 \langle B \rangle$ in the secular phase). At early times, the ion magnetic moment is well-conserved but once the fluctuation amplitudes exceed $\delta B \sim 0.1 \, \langle B \rangle$, the magnetic moment is no longer conserved but instead changes on a timescale comparable to that of the mean magnetic field. We discuss the implications of our results for low-collisionality astrophysical plasmas, including the near-Earth solar wind and low-luminosity accretion disks around black holes. \newline \end{abstract}

\keywords{accretion, accretion disks -- instabilities -- plasmas -- solar wind}

\section{Introduction}
\label{sec:intro}

\noindent Ion pressure anisotropies are ubiquitous in heliospheric and astrophysical plasmas. In the absence of Coulomb collisions, the magnetic moment of ions, $\mu_i$ ($\equiv v_{\perp,i}^2/B$, where $v_{\perp,i}$ is the ion velocity perpendicular to the local magnetic field $\vec{B}$ and $B=|\vec{B}|$) is an adiabatic invariant. Thus, if ion collisions are infrequent, the ion velocity distributions parallel and perpendicular to the magnetic field decouple, making $\Delta p_i \equiv p_{\perp,i} - p_{||,i} \ne 0$ (where $p_{\perp,i}$ and $p_{||,i}$ are the pressure components of ions perpendicular and parallel to $\vec{B}$). Examples of systems where ion pressure anisotropies are important are low-luminosity accretion disks around compact objects \citep{SharmaEtAl06}, the intracluster medium (ICM) \citep{SchekochihinEtAl05, Lyutikov07}, and the heliosphere \citep[see, e.g., ][]{hellinger2006, MatteiniEtAl07, MarucaEtAl11}. In these systems the ion pressure anisotropy is expected to play a key role in the large scale dynamics of the plasma.  For instance, \cite{SharmaEtAl06} pointed out that pressure anisotropies can give rise to an anisotropic stress that can contribute to angular momentum transport and plasma heating in low-luminosity accretion disks. This physics is not included in standard MHD models of accretion disks.\newline

\noindent Velocity space instabilities limit the growth of the pressure anisotropy and give rise to small scale fluctuations in the magnetic field. These fluctuations can, in turn, affect the large scale transport properties of the plasma by modifying the mean free path of particles.\newline 

\noindent Magnetic field amplification in a collisionless system generically drives $p_{\perp,i} > p_{||,i}$. There are two ion-scale instabilities that can be excited in this regime: the mirror and the ion-cyclotron (IC) instabilities \citep{Hasegawa69, Gary92,Southwood93}. The mirror instability consists of non-propagating, strongly compressional modes. Their fastest growing wave vectors $\vec{k}$ are oblique to $\vec{B}$, with the magnetic variations parallel to $\vec{B}$ being much larger than the perpendicular fluctuations, $\delta \vec{B}_{||} \gg \delta \vec{B}_{\perp}$. For a bi-Maxwellian distribution of ions, and assuming cold electrons, the threshold condition for mirror instability growth is given by $T_{\perp,i}/T_{||,i}-1 > 1/\beta_{\perp,i}$, where $T_{\perp,i}$ ($T_{||,i}$) is the ion temperature perpendicular (parallel) to $\vec{B}$, and $\beta_{\perp,i} \equiv 8\pi p_{\perp,i}/B^2$ \citep{Hasegawa69}. The IC instability, on the other hand, consists of transverse electromagnetic waves, with the fastest growing $\vec{k}$ being preferentially parallel to $\vec{B}$ \cite[e.g.,][]{AndersonEtAl91}. Whether the IC or mirror instability sets in first depends on how fast these instabilities grow for a given set of plasma conditions. 
\cite{GaryEtAl94} showed that for $T_{\perp,i}/T_{||,i}-1 = 0.35/\beta_{||,i}^{0.42}$, the growth rate of the IC instability is $\gamma_{IC} = 10^{-4}\omega_{c,i}$, where $\omega_{c,i}$ is the cyclotron frequency of the ions. In addition, for $\gamma_{IC}/\omega_{c,i} \ll 1$, the threshold anisotropy depends very weakly on $\gamma_{IC}$.
\newline

\noindent In the linear regime, the dominant instability will be determined by which threshold is reached first as $\vec{B}$ is amplified. These estimates for the threshold conditions imply that, in the linear regime, the IC instability should dominate for $\beta_i \sim 1$, while the mirror instability should dominate for $\beta_i \gg 1$.\newline

\noindent Although both the mirror and IC instabilities have been extensively studied in the linear regime, a complete theory of their nonlinear evolution and saturation is still lacking. In this paper, we are particularly interested in the question of how the mirror and IC modes behave after the initial phase of exponential growth. Indeed, in most astrophysical scenarios where the magnetic field is amplified, the growth of the field occurs on time scales much longer than the growth time of the relevant kinetic instabilities. Therefore, most of the evolution of the velocity space instabilities happens in a nonlinear regime, where the conditions typically assumed in linear studies, such as a homogeneous plasma or a bi-Maxwellian distribution of particle velocities, are not necessarily satisfied. Moreover, if the field is amplified by order unity or more, a quasi-linear analysis is not applicable. One of the key questions we are interested in addressing in this case is whether the mirror instability saturates via pitch-angle scattering that violates magnetic moment conservation \citep{SharmaEtAl06} or via a nearly $\mu_i$-conserving rearrangement of the magnetic field, with fluctuations $|\delta \vec{B}| \sim |\vec{B}|$ \citep{SchekochihinEtAl08}.  We shall see that both of these saturation processes can in fact be important.  \newline

\noindent The nonlinear regime of the IC and mirror instabilities has been studied both theoretically \citep[e.g.,][]{SchekochihinEtAl08,HellingerEtAl09} and numerically  \citep{BaumgartelEtAl03,TravnicekEtAl07,CalifanoEtAl08, GuoEtAl2009}. In most numerical studies, however, the pressure anisotropy is treated as an initial condition, which decreases as the instabilities grow and saturate. Thus, in these approaches the nonlinear saturation cannot be followed for a time much longer than that of the initial exponential growth.  Moreover, many of the previous numerical studies have focused on one-dimensional simulations.  \newline

\noindent In this paper we study the long term, nonlinear evolution of the mirror and IC instabilities using two-dimensional particle-in-cell (PIC) simulations in driven systems. In order to self-consistently explore a time much longer than the initial exponential growth, we continuously induce the growth of $\Delta p_i$ by amplifying the mean field $<\vec{B}>$ during the simulation. We concentrate on amplification by incompressible plasma motions. Thus,  our simulations impose a shear velocity on the plasma, which sustains the growth of the magnetic field and maintains an overall positive pressure anisotropy. This growth is intended to mimic magnetic field fluctuations in a turbulent plasma and MHD-scale instabilities that amplify the magnetic field, like the magnetorotational instability \citep[MRI;][]{BalbusEtAl91} or convective instabilities driven by anisotropic thermal conduction in dilute plasmas \citep{Balbus01,Quataert08}.  We suspect that our results are also relevant to cases in which perpendicular heating or parallel cooling drives a plasma mirror and/or IC unstable, as in, e.g., the solar wind.  \cite{HellingerEtAl08} conducted an analogous study to ours using hybrid simulations. In their study, expansion of the simulation box decreased the mean magnetic field, so that on average $p_{\perp,i} < p_{||,i}$. This in turn led to the growth and nonlinear saturation of the firehose instability.  Also, \cite{TravnicekEtAl07} conducted a similar hybrid simulation study where the simulation box is expanded along the directions parallel and perpendicular to $\vec{B}$. With this setup, $p_{\perp,i}$ becomes larger than $p_{||,i}$ on average, and both the mirror and IC instabilities can grow and reach saturation. Our work is complementary to these hybrid simulation studies and focuses on the process in which the condition $p_{\perp,i} > p_{||,i}$ is achieved by field amplification due to shearing plasmas. We also consider higher values of $\beta_i$ $\approx 1-80$, relevant to accretion disks, the ICM, and the near-Earth solar wind. Our calculations are less directly applicable to heliospheric measurements of mirror modes driven unstable by rapid heating through the Earth's bow shock \citep[e.g.,][]{SchwartzEtAl96}. This is because we assume that the pressure anisotropy is generated on a timescale long compared to the ion cyclotron timescale, which is probably not true at collisionless shocks.\newline

\noindent Our paper is organized as follows. In \S \ref{sec:numsetup} we describe the simulation set up, emphasizing the key physical and numerical parameters. \S \ref{sec:results} shows our results and \S \ref{sec:conclu} presents our conclusions.  Throughout the paper we compare some of our simulation results with linear theory predictions appropriate to our PIC simulations.  These have artificially low ion to electron mass ratios ($m_i/m_e \simeq 1-10)$.  The linear theory results are based on the linear Vlasov solver developed in \citet{Verscharen2013}.\newline

\noindent During the completion of this work, \citet{KunzEtAl14} presented calculations of firehose and mirror saturation in shearing plasmas with very similar results to those that we present here.\newline

\section{Simulation Setup}
 \label{sec:numsetup}
 We use the electromagnetic, relativistic PIC code TRISTAN-MP \citep{Buneman93, Spitkovsky05} in two dimensions. The simulation box consists of a square box in the $x-y$ plane, containing plasma with a homogeneous magnetic field $\vec{B}_0$ initially pointing along the $\hat{x}$ axis. Since we want to simulate a magnetic field that is being amplified in an incompressible way, we impose a shear motion of the plasma so that the mean particle velocity is $\vec{v} = -sx\hat{y}$, where $s$ is a shear parameter with units of frequency and $x$ is the distance along $\hat{x}$. From flux conservation, the $y$-component of the mean field evolves as $\partial <B>_y/\partial t = -sB_0$, implying a net growth of $|<\vec{B}>|$. Although we present simulations resolving the $x-y$ plane only, we also tried runs where the $x-z$ plane was resolved. In those cases the isotropization efficiency was substantially lower. This is because, if the $x-z$ plane is resolved, the growing component of $\vec{B}$ is perpendicular to the plane of the simulation. This way the angle between the relevant wave vectors, $\vec{k}$, and $\vec{B}$ cannot be 0. This over-constrains the wave vectors that are allowed to exist, artificially reducing the isotropization efficiency.\newline 

\noindent An important physical parameter in our simulations will be the ratio of the initial ion cyclotron frequency to the shear frequency, $\omega_{c,i}/s$. We refer to this as the magnetization.  In typical astrophysical environments $\omega_{c,i} \gg s$. Due to computational constraints, however, we will use values of $\omega_{c,i}/s$ much smaller than expected in reality, although still satisfying $\omega_{c,i}/s \gg 1$. The dependence of our results on the ratio $\omega_{c,i}/s$ will be carefully assessed.\newline

\begin{deluxetable*}{ccccccccc} \tablecaption{Physical and numerical parameters for the runs} \tablehead{ \colhead{Runs}&\colhead{$m_i/m_e$}&\colhead{$\beta_{init}$}&\colhead{$\beta_{init,i}$ ($=\beta_{init,e}$)}&\colhead{$\omega_{c,i}/s$}&\colhead{$v_{A,0}/c$}&\colhead{$c/\omega_{p,e}/\Delta_x$}&\colhead{N$_{ppc}$}&\colhead{L/R$_{L,i}$} } \startdata
  beta6mag93 &  1&  6 & - & 93 & 0.05&  14 & 30 & 12 \\
  beta6mag670a &  1&  6& - & 670 & 0.15&  14 & 10 & 18 \\
  beta6mag670b &  1&  6& - & 670 & 0.15&  14 & 10 & 24 \\
  beta20mag93a &  1&  20& - & 93 & 0.05&  10 & 60 & 35 \\
  beta20mag93b &  1&  20& - & 93 & 0.05&  10 & 30 & 35 \\
  beta20mag670a &  1&  20& - & 670 & 0.05&  7 & 30 & 36 \\
  beta20mag670b &  1&  20& - & 670 & 0.05&  7 & 30 & 48 \\
  beta20mag2000 &  1&  20& - & 2000 & 0.05&  7 & 50 & 55 \\
  beta80mag670 &  1&  80& - & 670 & 0.05&  5 & 30 & 24 \\
  beta80mag1340 &  1&  80& - & 1340 & 0.05&  5 & 30 & 24 \\
  betai20magi240mass10 & 10& -& 20 & 240 & 0.05& 5 & 20 & 25 \enddata \tablecomments{A summary of the physical and numerical parameters of the simulations discussed in the paper. These are the mass ratio $m_i/m_e$, $\beta_{init}$ ($=\beta_{init,i} + \beta_{init,e}$) if $m_i/m_e=1$, $\beta_{init,i}$ ($=\beta_{init,e}$) if $m_i/m_e > 1$, the magnetization $\omega_{c,i}/s$, the initial Alfv\'en velocity, $v_{A,0}$, the skin depth $c/\omega_{p,e}/\Delta_x$ (where $\Delta_x$ is the grid points separation), the number of particles per cell N$_{ppc}$ (including ions and electrons), and the box size in units of the typical ion Larmor radius $L/R_{L,i}$ ($R_{L,i} = v_{th,i}/\omega_{c,i}$, where $v_{th,i}^2=3p_i/\rho$ is the rms ion velocity and $\rho$ is the mass density of the ions). We confirmed numerical convergence by exploring the resolution in $c/\omega_{c,e}/\Delta_x$, N$_{ppc}$, and L/R$_{L,i}$ for all parameter combinations of $\beta_{init}$ (or $\beta_{init,i}$), $\omega_{c,i}/s$, and $m_i/m_e$.} \label{table:1D} \end{deluxetable*} 

\noindent In standard MHD simulations where shear plasma motions are imposed, shearing periodic boundary conditions would be used along the $x$ direction \citep[see, e.g.,][]{HawleyEtAl95}. In that case, 
the flow velocities at the $x$-boundaries of the box are matched using Galilean transformations of the fluid velocity. In the case of relativistic PIC simulations this can not be done in a self-consistent way. The reason is that, under a relativistic change of reference frame, the current $\vec{J}$ transforms in a way that is inconsistent with simply transforming the particle velocities.\footnote{Relativistic velocity transformations imply changes in the volume occupied by the particles, leading to a non-conservation of charge density that can not be trivially implemented in a PIC simulation \citep[see][]{RiquelmeEtAl12}.} We instead implement {\it shearing coordinates}, in which the grid moves with the shearing velocity $\vec{v}=-xs\hat{y}$. In this new frame the average plasma velocity vanishes everywhere in the box and simple periodic boundary conditions are allowed both in the $x$ and $y$ axes. In these shearing coordinates, Maxwell's equations gain additional terms \citep[see Appendix A of ][]{RiquelmeEtAl12}, becoming 
\begin{eqnarray}
\frac{\partial \vec{B}(\vec{r},t)}{\partial t} &= &-c\nabla\times\vec{E}(\vec{r},t) - sB_x(\vec{r},t)\hat{y} + \nonumber \\
&&s\Big(ct \frac{\partial \vec{E}(\vec{r},t)}{\partial y} + \frac{y}{c}\frac{\partial \vec{E}(\vec{r},t)}{\partial t}\Big) \times \hat{x} \textrm{   } \textrm{  } \textrm{    and} 
\label{eq:ind}
\end{eqnarray}
\begin{eqnarray}
\frac{\partial \vec{E}(\vec{r},t)}{\partial t} &=& c\nabla\times\vec{B}(\vec{r},t) -4\pi \vec{J} - sE_x(\vec{r},t)\hat{y} - \nonumber \\
&&s\Big( ct \frac{\partial \vec{B}(\vec{r},t)}{\partial y} + \frac{y}{c} \frac{\partial \vec{B}(\vec{r},t)}{\partial t} \Big)\times \hat{x},
\label{eq:amp}
\end{eqnarray}
where $c$ is the speed of light, and $\vec{E}$ is the electric field. In addition to the modifications to Maxwell's equations, the forces on the particles also acquire an extra term:
\begin{equation}
\frac{d\vec{p}}{dt} = s p_x\hat{y} +q(\vec{E}+\frac{\vec{u}}{c}\times\vec{B}),
\label{eq:fuerza}
\end{equation} 
where $\vec{p}$, $\vec{u}$, and $q$ are the particle's momentum, velocity, and charge, respectively. The third term on the right hand side of Equation \ref{eq:ind} and the fourth term on the right hand side of Equation \ref{eq:amp} are proportional to time and arise from the motion of the shearing coordinate grid points with respect to those of the non-shearing coordinates. Indeed, as time goes on, the $x$ axis of the shearing coordinates is gradually tilted (as seen from the non-shearing frame); therefore the $x-$derivatives of the fields ($\partial/\partial x$) in the shearing coordinates must include a time dependent term that accounts for the evolution of the $x$ axis.\newline
\begin{figure*}[t]
\centering
\includegraphics[width=18cm]{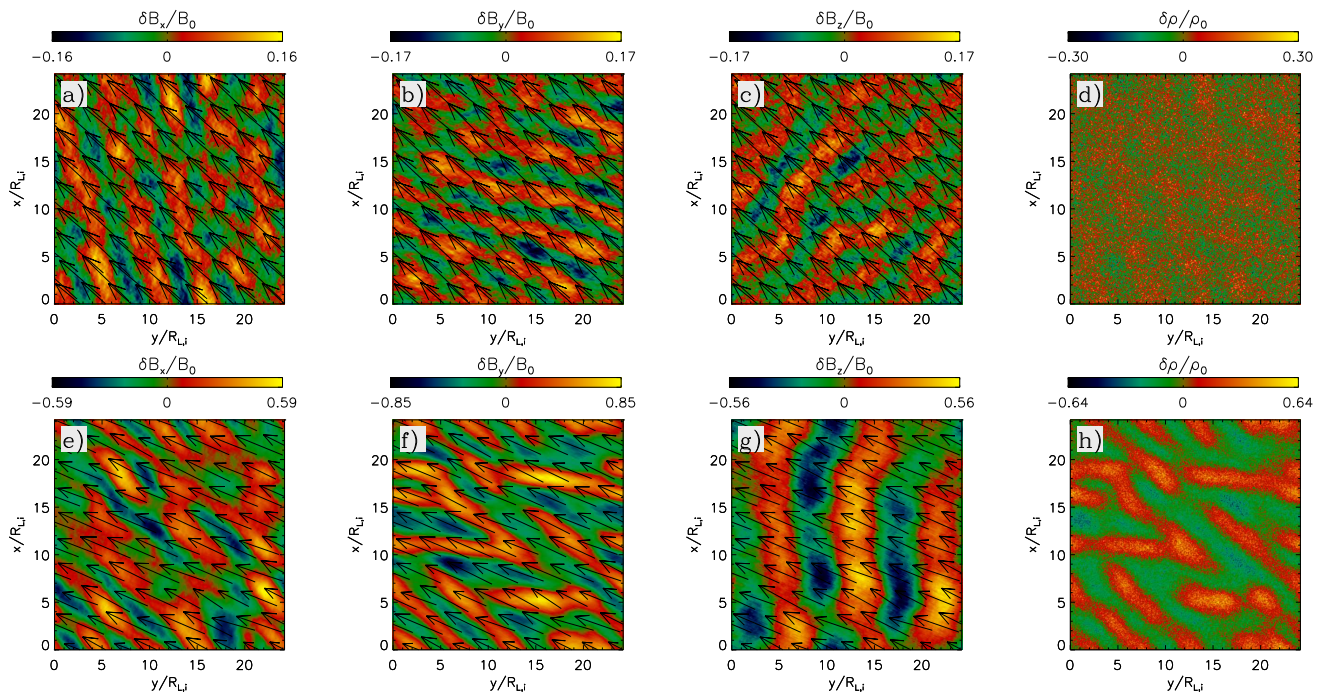} \caption{The three components of $\delta \vec{B}$ and plasma density fluctuations $\delta \rho$ at two different times: $t \cdot s=1$ (upper row) and $t \cdot s = 2$ (lower row), for run beta6mag670b ($\beta_{init}=6$,  $\omega_{c,i}/s=$670, $m_i/m_e=1$). Fields and density are normalized by $B_0$ and the initial density $\rho_0$, respectively. Arrows denote the mean magnetic field direction on the simulation plane. IC and mirror instabilities contribute comparably to the fluctuations at $t \cdot s=1$, while mirror dominates at $t \cdot s=2$. Density fluctuations best correlate with the mirror modes. The dominance of mirror fluctuations suggests that mirror modes are more robust than the IC modes in the saturated regime. This may be due to the particle energy spectrum departing from bi-Maxwellian (see Figure \ref{fig:spectrabeta6}), suppressing the growth of the IC modes.}  \label{fig:fieldsbeta6} \end{figure*}
\noindent The last terms in Equations \ref{eq:ind} and \ref{eq:amp}, which are proportional to the coordinate $y$, will be neglected for the following reasons. In our simulations the box size will typically be a few times the Larmor radius of the ions, $R_{L,i}$. Thus, $sy/c \sim (s/\omega_{c,i})(v_{th,i}/c)$ (where $v_{th,i}$ is the thermal velocity of the ions), which is much smaller than unity since we are interested in the regime $s/\omega_{c,i} \ll 1$ and $v_{th,i}/c \ll 1$. As a result, the last term in Equation \ref{eq:ind} will be much smaller than the term on the left hand side, especially since $|\vec{B}| \gg |\vec{E}|$.\newline 

\noindent The last term of Equation \ref{eq:amp}, on the other hand, is not necessarily much smaller than the displacement current (left hand side term), mainly because we expect $|\vec{E}| \ll |\vec{B}|$. However, if the characteristic timescale for the mirror and IC modes is close to $s^{-1}$ (below we will check that this is indeed the case), the last term in Equation \ref{eq:amp} will be much smaller than the first term on the right hand side (the ratio between these terms scales as $\sim (s/\omega_{c,i})^2(v_{th,i}/c)^2$), so we choose to neglect it in the limit $s \ll \omega_{c,i}$. The fact that the neglected term can still be comparable to the displacement current implies that we may be excluding a relevant contribution to the charge density in Gauss' law. However, the physics of interest in this paper is non-relativistic and charge separation does not play a role.  As a result, $c\nabla \times \vec{B} + sct\partial \vec{B}/\partial y \times \hat{x} \approx 4\pi \vec{J}$ (equivalent to $c\nabla \times \vec{B} \approx 4\pi \vec{J}$ in the {\it non-shearing} coordinate system). Neglecting the terms proportional to $y$ is also required by the periodic boundary conditions in the y-direction. The existence of these terms at all is a consequence of the incompatibility of the Galilean invariance of the shearing box and the Lorentz invariance of the PIC calculations. For consistency we will also neglect the term $sE_x\hat{y}$ in Equation \ref{eq:amp}, which is also comparable to the displacement current (in the case where $\vec{E}$ evolves on a timescale close to $s^{-1}$).\footnote{The modification to Faraday's equation that depends on $B_x$ can be integrated using simple time and space interpolations of $B_x$. This way, after this modification is implemented, the numerical algorithm used by TRISTAN-MP continues to be second order accurate in time and space.} 
\newline

\noindent Our set of simulations are summarized in Table 1. The physical setup is defined by the magnetization parameter, $\omega_{c,i}/s$ (the ratio of the initial cyclotron frequency to the shear rate), the ratio between the initial particle and magnetic pressures, $\beta_{init}=8\pi p_{init}/B_0^2$, the mass ratio between ions and electrons, $m_i/m_e$, and the initial Alfv\'en velocity of the plasma $v_{A,0}$, relative to the speed of light $c$, where $v_{A,0}\equiv B_0/\sqrt{4\pi \rho}$ and $\rho$ is the mass density. The numerical parameters of our runs are defined by the spatial resolution ($c/\omega_{p,e}/\Delta_x$), box size relative to the ion Larmor radius ($L/R_{L,i}$), and number of particles per cell ($N_{ppc}$), where $\omega_{p,e}$ is the plasma frequency of the electrons and $\Delta x$ is the spacial separation of the grid points. The simulations shown in this paper are described in Table 1. These simulations are a subset of a higher number of runs, where we used different combinations of numerical and physical parameters, which confirmed numerical convergence.\newline
\vspace{0.3cm}
\section{Results}
\label{sec:results}

\noindent We want to explore the regime $\beta_{\perp}, \beta_{||} \approx 1-80$. In order to do so, most of our runs use the mass ratio $m_i/m_e=1$. Since for $m_i/m_e=1$ both species will be essentially indistinguishable, we initially give ions and electrons Maxwellian energy distributions with the same temperature, and use the parameters $\beta_{\perp}=\beta_{e,\perp} + \beta_{i,\perp}$ and $\beta_{||}=\beta_{e,||} + \beta_{i,||}$ to quantify the pressure of the plasma perpendicular and parallel to $\vec{B}$. Simulations with $m_i/m_e=1$ of course do not allow us to study the physics of electron isotropization. However, in \S \ref{sec:compmime} we use an example of our $m_i/m_e>1$ runs to show that $m_i/m_e=1$ calculations reproduce the ion-related phenomena fairly well, with $\beta_{\perp}$ and  $\beta_{||}$ playing the same role of $\beta_{i,\perp}$ and $\beta_{i,||}$ in the $m_i/m_e>1$ runs. We defer a detailed study of electron scale pressure isotropization to future work.\newline 

\noindent We divide our runs into cases with three different initial betas: $\beta_{init}=6$, 20, and 80. Our analysis focuses on the nonlinear structure of the IC/mirror generated fluctuations, the evolution of the pressure anisotropy $p_{\perp}/p_{||}-1$, and the conservation of the ion magnetic moment.

\subsection{Case $\beta_{init}=6$}
\begin{figure}
\centering
\includegraphics[width=9cm]{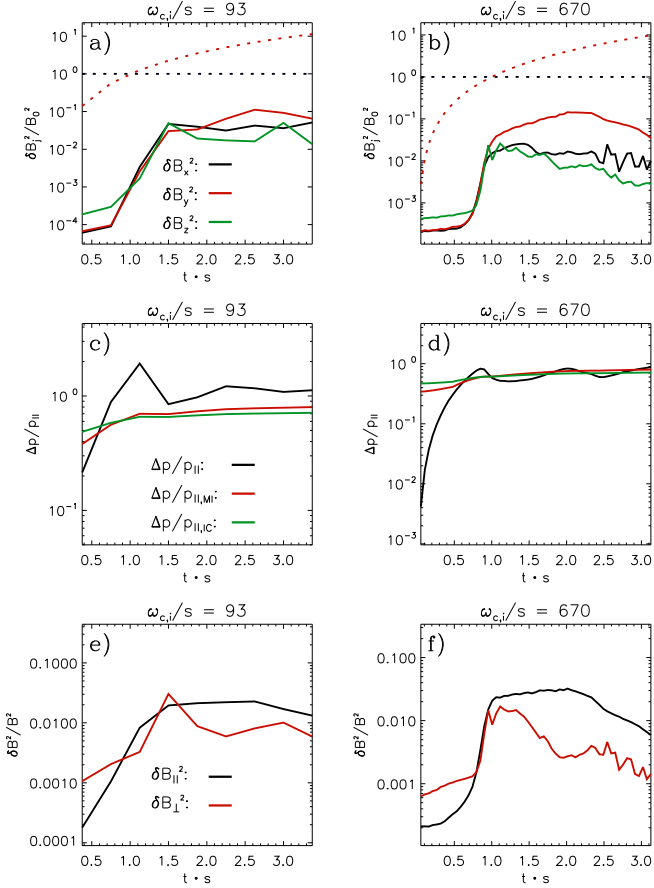}
\caption{Field fluctuations and pressure anisotropy for our $\beta_{init} = 6$ runs.  Panels $a$ and $b$ show the evolution of $\delta B_j^2/B_0^2$ ($\equiv <(B_j-<B_j>)^2>/B_0^2$; solid) and $B_j^2/B_0^2$ ($\equiv <B_j>^2/B_0^2$; dotted) for runs beta6mag93 and beta6mag670a, respectively ($j=x, y,$ and $z$ correspond to black, red, and green, respectively). For the same runs, panels $c$ and $d$ show $\Delta p/p_{||}$ ($\equiv <(p_{\perp}-p_{||})/p_{||}>$; black-solid), compared with the linear mirror and IC thresholds  ($\Delta p/p_{||,MI}$ and $\Delta p/p_{||,IC}$, in red and green, respectively) for pair plasma and growth rates $\gamma=0.05 \omega_{c,i}$ and $\gamma=0.007 \omega_{c,i}$, respectively. Panels $e$ and $f$ show $\delta B_{||}^2/B^2$ 
and $\delta B_{\perp}^2/B^2$,
where the subscripts $||$ and $\perp$ denote the components parallel and perpendicular to $<\vec{B}>$, respectively.   After the plasma pressure anisotropy exceeds the mirror and IC thresholds, there is an initial phase of exponential growth followed by a secular phase.   The pressure anisotropy initially grows but saturates at the linear threshold for the mirror/IC instabilities.}
\label{fig:beta6}
\end{figure}

\begin{figure}
\centering
\includegraphics[width=9cm]{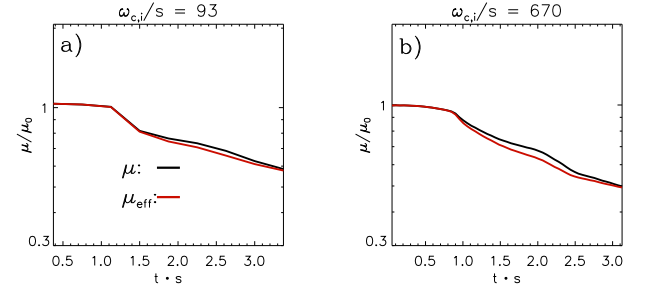} \caption{Evolution of the ion magnetic moment for the $\beta_{init} = 6$ runs with $\omega_{c,i}/s = 93$ and $670$ (beta6mag93 and beta6mag670a in Table \ref{table:1D}).  We show the evolution of the true average magnetic moment $\mu$ $\equiv <v_{\perp}^2/B>_p$ (black line) and an effective global magnetic moment, $\mu_{\rm eff}$ $\equiv <v_{\perp}^2>_p/|<\vec{B}>|$ (red line), where the subscript $p$ denotes an average over all particles.  The ion magnetic moment $\mu$ is conserved until $t s \sim 1$, at which point it decreases on the same timescale as the mean magnetic field (see Fig. \ref{fig:muderivs}).   
The close similarity between $\mu$ and $\mu_{\rm eff}$ shows that there are not strong correlations between the particle velocity $v_\perp$ and magnetic field $B$; such correlations are more prominent at higher $\beta_{init}$ where the IC instability is subdominant (see Fig. \ref{fig:mubeta20}).}
\label{fig:beta6b}
\end{figure}

\begin{figure}
  \centering \includegraphics[width=7.5cm]{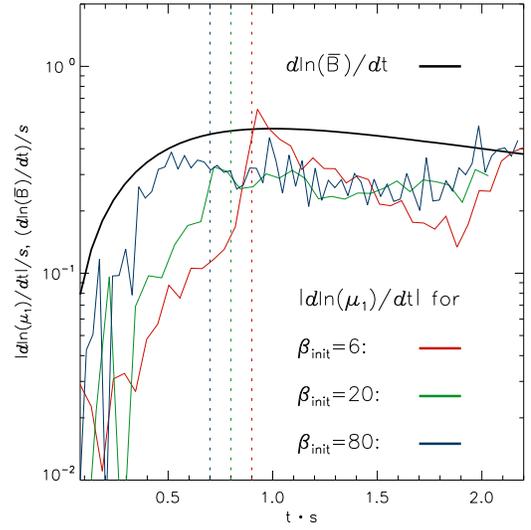}
  \caption{The rate at which the ion magnetic moment changes in time, $|d\ln{\mu}/dt|$, for different $\beta_{init}$, compared to the growth rate of the mean magnetic field $d\ln{\langle |{\vec B}| \rangle}/dt$ (black line; this quantity is the same for the different $\beta_{init}$ runs).  $|d\ln{\mu}/dt|$ is shown for our simulations with $\beta_{init} = 6, 20,$ and 80 (beta6mag670a, beta20mag670a, and beta80mag1340, respectively). The vertical-dotted lines mark the beginning of the saturated state for each run based on when the growth of the fluctuations becomes roughly secular.  In the saturated state, the magnetic moment changes on a timescale comparable to that of the mean magnetic field (to within $\sim 25-50 \%$).  At earlier times, however, the magnetic moment is reasonably well conserved.}
\label{fig:muderivs}
\end{figure}

\noindent As discussed above, the expectation is that the IC instability should play an important role in the $\beta_{init}=6$ case, and it should become significantly less important in the $\beta_{init}=20$ and 80 cases. 
\noindent  The contributions from the IC and mirror modes can be seen from Figure \ref{fig:fieldsbeta6}. This figure shows 2D images of the field fluctuations $\delta B_j/B_0$ ($\equiv (B_j-<B_j>)/B_0$; $< >$ stands for ``volume average") and the plasma density fluctuations $\delta \rho/\rho_0$ for run beta6mag670$c$ (``$j$" stands for the component $x$, $y$, or $z$ of $\delta \vec{B}$) at two different times: $t\cdot s=1$ (upper panels) and $t\cdot s=2$ (lower panels).\newline 

\noindent We can see that at the earlier time ($t\cdot s=1$) the three components of $\delta \vec{B}$ have about the same amplitude (see Figures \ref{fig:fieldsbeta6}$a$, \ref{fig:fieldsbeta6}$b$, and \ref{fig:fieldsbeta6}$c$), but appear to be dominated by different mode orientations, indicating the simultaneous presence of IC and mirror modes. The mirror modes are expected to have $\delta \vec{B}$ mainly in the plane of the simulation \citep[the plane of $\vec{k}$ and $<\vec{B}>$;][]{PokhotelovEtAl04}. On the other hand, the IC modes have the three components of $\delta \vec{B}$ of comparable magnitude, so their presence can be most clearly revealed by $\delta B_z$ (Figure \ref{fig:fieldsbeta6}$c$). Indeed, while $\delta B_x$ and $\delta B_y$ are dominated by a combination of oblique (mirror) and quasi-parallel (IC) modes, $\delta B_z$ appears to be dominated by only quasi-parallel modes. Thus, initially the mirror and IC instabilities contribute comparably to $\delta \vec{B}$. At the later time ($t\cdot s=2$), however, the fluctuations become dominated by an oblique wave vector, with the IC instability playing a subdominant role. The density fluctuations at all times are $\delta \rho/\rho_0 \ll 1$, and seem to correlate primarily with the mirror modes.\newline

\noindent The relative contribution of the mirror and IC instabilities can also be seen from Figure \ref{fig:beta6}, which shows the time evolution of different volume-averaged quantities for runs with $\omega_{c,i}/s=93$ (first column; run beta6mag93$a$) and $\omega_{c,i}/s=670$ (second column; run beta6mag670$a$). In Figures \ref{fig:beta6}$a$ and \ref{fig:beta6}$b$ we plot $\delta B_j^2/B_0^2$ for the two magnetizations. In both cases there is an initially exponential growth of $|\delta \vec{B}|$ until $<\delta B_j^2>/B_0^2 \sim 0.03$. Both instabilities have about the same growth rates, which are $\gamma_{IC}\approx \gamma_{MI} \sim 10 s$. Note that the dominant growth rate of the mirror/IC modes is nearly the same in the two simulations with different values of $\omega_{c,i}$.  The growth rate is thus set by the background shear rate $s$ instead of $\omega_{c,i}$. This can be understood by noting that the smaller growth rate modes have a smaller anisotropy threshold. Therefore, as $B$ and $\Delta p$ grow, modes with smaller growth rate will begin to grow first. This implies that the dominant modes will be those that can reach a significant amplitude to stop the growth of $\Delta p$ on the shear timescale $s^{-1}$. \newline

\begin{figure}
\centering
\includegraphics[width=9cm]{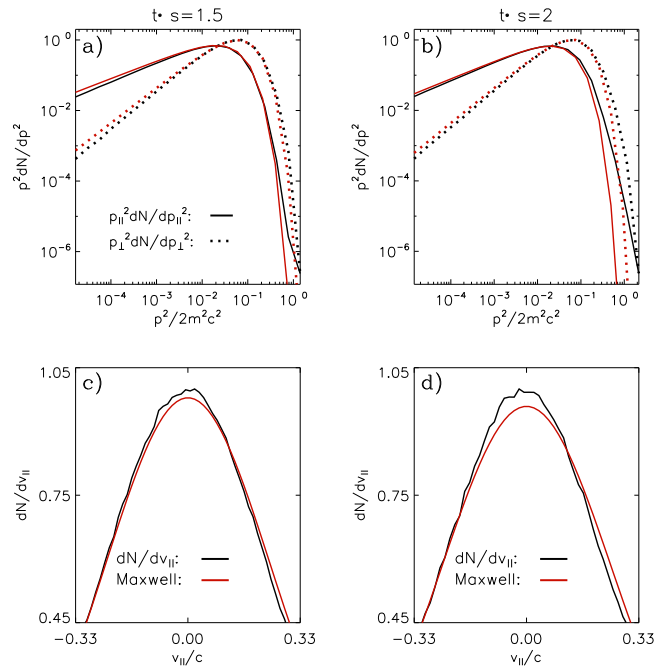}
\caption{The energy spectra of particles for the high magnetization $\beta_{init} = 6$ run beta6mag670a at two different times.  {\em Top:}  Energy spectra as a function of the energies perpendicular and parallel to the magnetic field, $p_{\perp}^2/2m$ (black-dotted line) and $p_{||}^2/2m$ (black-solid line).  {\em Bottom:}   Energy spectra as a function of $v_\parallel$.  
Bi-Maxwellian fits are shown in red. There is a clear deviation from the bi-Maxwellian at the highest energies, which grows in time.   There is, however, little deviation from a Maxwellian at $v_\parallel \sim 0$, in contrast to some of the 1D runs discussed in the Appendix.  Comparison with Figure \ref{fig:spectrabeta20} suggests that the high energy deviation from Maxwellian is due to the IC instability, since it is less prominent at higher $\beta$ when the IC instability is sub-dominant.}
\label{fig:spectrabeta6}
\end{figure}

\noindent After the exponential phase, the growth becomes dominated by the mirror modes and is secular \citep[as predicted by][]{SchekochihinEtAl08}.  This transition from exponential to secular  is clearly seen at $t\,s \sim 1$ for $\omega_{c,i}/s = 670$ in Figure \ref{fig:beta6}. Despite the similar exponential growth rate for the mirror and IC instabilities in the linear regime, the mirror modes are more robust and dominate in the nonlinear regime. This dominance of the mirror modes is remarkable because it occurs even for $\beta_i \sim 1$, where the linear analysis predicts an important contribution of the IC instability (in the nonlinear regime the mean magnetic energy has been amplified by about one order of magnitude, so that $\beta_i \sim 1$). We suggest below (Fig. \ref{fig:spectrabeta6}) that departure from the bi-Maxwellian energy distribution assumed in the linear instability analysis is likely playing an important role in this sub-dominance of the IC instability.\newline

\noindent One difference in the evolution of $\delta \vec{B}$ for the two different magnetizations is that the end of the exponential phase occurs at somewhat higher amplitude in the $\omega_{c,i}/s=93$ case. This difference implies that, for smaller values of  $\omega_{c,i}/s$, the mirror and IC fluctuations are less efficient in suppressing the pressure anisotropy, requiring larger values of $|\delta \vec{B}|$. At the end of this section, we will explain in further detail this dependence of $|\delta \vec{B}|$ on magnetization by focusing on the mechanism by which the isotropization occurs.\newline

\noindent One consequence of the difference in $|\delta \vec{B}|$ is that the lower magnetization runs take longer to reach the point when $\Delta p/p_{||}$ ($\equiv <(p_{\perp}-p_{||})/p_{||}>$) saturates. Thus in the $\omega_{c,i}/s=93$ run the anisotropy $\Delta p/p_{||}$ has more time to grow, reaching larger values before saturating. This can be seen in the evolution of $\Delta p/p_{||}$, shown in black lines in Figures \ref{fig:beta6}$c$ and \ref{fig:beta6}$d$ for runs beta6mag93$a$ and beta6mag670$a$, respectively.  For both magnetizations, there is an initial overshoot in $\Delta p/p_{||}$, whose amplitude is larger for the $\omega_{c,i}/s=93$ run. After the overshoot, $\Delta p/p_{||}$ evolves in a similar way for both magnetizations. 
\newline

\noindent Since in the nonlinear regime $\delta \vec{B}$ is dominated by mirror modes, after the saturation one would expect the pressure anisotropies to behave according to the marginal stability conditions of the mirror instability.
Since in our simulations both the IC and mirror modes grow at $\gamma \sim 10 s$, we calculated the IC and mirror thresholds for equivalent growth rates in the pair plasma case with magnetizations $\omega_{c,i}/s=93$ and $\omega_{c,i}/s=670$. Thus, Figures \ref{fig:beta6}$c$ and \ref{fig:beta6}$d$ show that the threshold for mirror (red) and IC instability (green) in the cases $\gamma = 0.05\omega_{c,i}$ and $\gamma = 0.007\omega_{c,i}$, respectively. We see that the mirror threshold coincides fairly well with the saturated $\Delta p/p_{||}$ for both magnetizations. These results imply that the linear marginal stability condition for the mirror instability is fairly accurate in determining $\Delta p/p_{||}$ in the nonlinear regime. In realistic astrophysical setups, where $\omega_{c,i}$ can be many orders of magnitude larger than $s$, the expectation is that $\Delta p/p_{||}$ will follow the $\gamma \to 0$ mirror threshold, given by $p_{\perp}/p_{||}-1 = 1/\beta_{\perp}$. This differs from the results shown in Figure \ref{fig:beta6}, which are only appropriate for pair plasmas and modest magnetization.\newline \begin{figure}
  \centering \includegraphics[width=7.5cm]{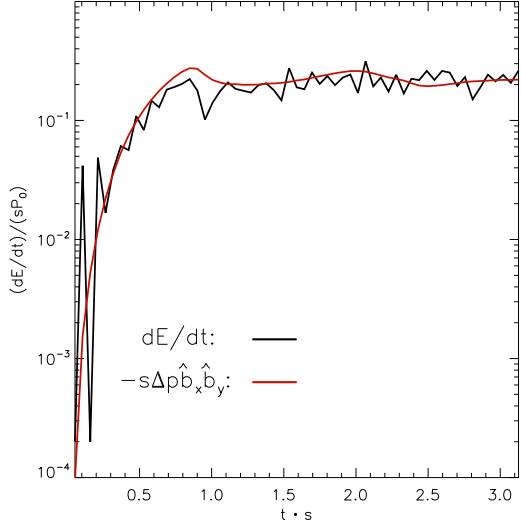} \caption{The time derivative of the total particle thermal energy (black) along with the volume averaged heating rate by the anisotropic stress $-s\Delta p\hat{b}_x\hat{b}_y$ (red) for $\beta_{init} = 6$ and $\omega_{c,i}/s = 670$ (run beta6mag670a).  The good agreement between the two results implies that the anisotropic stress coupling to the background shear is the primary mechanism for particle heating.\newline}  \label{fig:anistress} \end{figure}

\noindent As discussed in the introduction, it is not known whether the mirror instability saturates via pitch-angle scattering \citep[see, for instance,][]{SharmaEtAl06} or whether it cancels the appearance of a pressure anisotropy by substantially modifying the structure of the magnetic field \citep{SchekochihinEtAl08}. In the latter case $\delta \vec{B}$ should grow secularly until $|\delta \vec{B}|\simeq B$, with the breaking of $\mu-$invariance not necessary for the regulation of $p_{\perp}$. Figures \ref{fig:beta6}$e$ and \ref{fig:beta6}$f$ show the evolution of $\delta B_{\perp}^2/B^2$ 
and $\delta B_{||}^2/B^2$ 
for $\omega_{c,i}/s = 93$ and $670$, respectively, 
where $\vec{B}_{\perp}$ and $\vec{B}_{||}$ refer to the magnetic field component perpendicular and parallel to the mean field $<\vec{B}>$. We see that $i)$ $\delta B_{||}^2$ dominates in the nonlinear regime, as expected from the larger amplitude of the oblique mirror modes, and $ii)$ the saturation amplitude is  $|\delta \vec{B}|^2/B^2 \sim 0.04$, which favors the scenario where the non-conservation of $\mu$ is the key mechanism for the isotropization of the plasma pressure, at least if the background field has been amplified significantly.   \newline

\noindent To quantify the time variation of the average ion magnetic moment we define
\begin{equation}
\mu = \left<\frac{v_{\perp}^2}{B}\right>_p \textrm{ }\textrm{ and }\textrm{ } \mu_{\rm eff} = \frac{<v_{\perp}^2>_p}{|<\vec{B}>|}, 
\label{eq:mu}
\end{equation}
where $< >_p$ stands for average over all the particles. $\mu$ corresponds to the actual average of the magnetic moment.  Thus if a nearly $\mu-$conserving process dominates the mirror saturation, $\mu$ should remain fairly constant.  $\mu_{\rm eff}$, on the other hand, can be thought as an {\it effective} global magnetic moment that ignores the fluctuations in ${\vec B}$ or the correlation between $v_\perp$ and $B$ due to particles collecting in mirrors.  $\mu_{\rm eff}$ necessarily decreases since the mirror and IC instabilities suppress the growth that $<v_{\perp}^2>_p$ would have if the particles were only affected by the mean field $<\vec{B}>$.  Thus, if $\mu$ is nearly conserved, $\mu$ and $\mu_{\rm eff}$ should evolve quite differently, with $\mu > \mu_{\rm eff}$.  On the other hand, if there is significant pitch angle scattering $\mu \sim \mu_{\rm eff}$ and both should decrease on the same timescale that $\langle {\vec B} \rangle$ increases.\newline

\noindent Figures \ref{fig:beta6b}$a$ and \ref{fig:beta6b}$b$ show $\mu$ and $\mu_{\rm eff}$ for our $\beta_{init} = 6$ runs with magnetizations $\omega_{c,i}/s=93$ and $\omega_{c,i}/s=670$, respectively.
We see that for both magnetizations the difference between $\mu$ and $\mu_{\rm eff}$ is very small (only $\sim$ a few $\%$ difference). This implies that there is little spatial correlation between $v_\perp$ and $B$.  We shall see below that $\mu$ and $\mu_{\rm eff}$ differ somewhat more at higher $\beta$ where the mirror instability dominates over the IC instability.
\newline

\noindent  Figure \ref{fig:muderivs} compares the rate of change of the ion magnetic moment with that of the mean magnetic field for runs with $\beta_{init}=$6, 20, and 80.  This comparison is important since it is the evolution of the mean field that is driving the velocity space instabilities in our calculations.  
If the total thermal energy varies on a timescale long compared to the mean magnetic field, maintaining marginal stability to the mirror instability via pitch-angle scattering implies $|d \ln\mu/dt| \simeq |d \ln B/dt|[1 - \mathcal{O}(\beta^{-1})]$. The term $ \mathcal{O}(\beta^{-1})$ is exactly $\beta^{-1}$ for the canonical high magnetization mirror instability threshold but is somewhat different in our pair plasma, modest magnetization simulations (hence the use of $\mathcal{O}$).  This expression for $d \ln \mu/dt$ implies that pitch-angle scattering should lead to the magnetic moment varying at a rate slightly less than that of the mean magnetic field.  This is consistent with our numerical results in Figure \ref{fig:muderivs} in the saturated state where the magnetic field grows secularly in time.  \newline

\noindent At earlier times, when the magnetic field fluctuations due to the mirror and IC instability are smaller, $\mu$ is approximately conserved even though the mirror and IC are present and grow exponentially.  In particular, there is a temporal lag of $\sim 0.3 s^{-1}$ between the onset of the exponential growth of the mirror and IC instabilities and the onset of pitch angle scattering that decreases $\mu$.  Our interpretation of this result is that the mirror/IC fluctuations must reach a sufficient amplitude, roughly $\delta B \sim 0.1-0.3 \, \langle B \rangle$, in order for pitch angle scattering to be effective.  For the mirror instability, this violation of $\mu$ conservation by finite amplitude fluctuations is likely due to the stochasticity of particle orbits that sets in for large amplitude (albeit low frequency) fluctuations (e.g., \citealt{ChenEtAl01, JohnsonEtAl01}).  For the ion cyclotron instability, the need for finite amplitudes before significant scattering sets in is a consequence of the well-known result that the scattering rate for cyclotron-frequency fluctuations is $\sim \omega_{c,i} (\delta B/B)^2$.\newline

\begin{figure*}
\centering
\includegraphics[width=13cm]{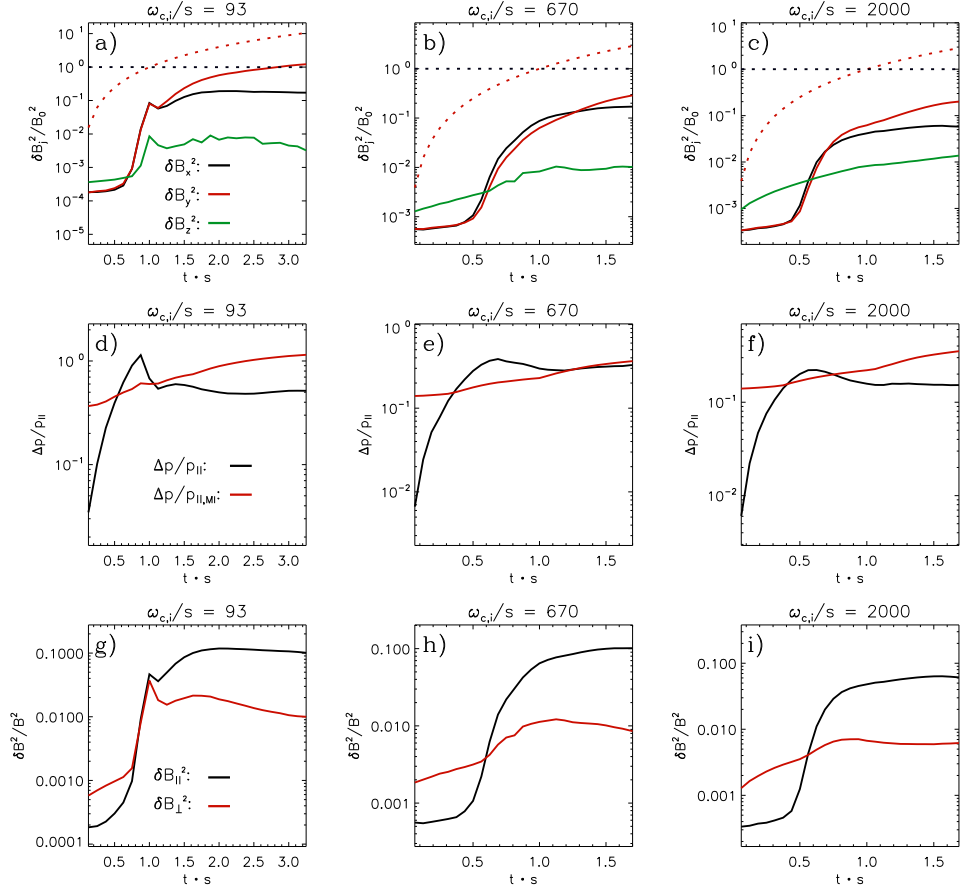}
\caption{Magnetic field fluctuations and pressure anisotropy for our $\beta_{init} = 20$ runs with $\omega_{c,i}/s = 93, 670, \& \, 2000$ (runs beta20mag93a, beta20mag670a, and  beta20mag2000).   The results for $\omega_{c,i}/s = 670$ and $2000$ are quantitatively similar, indicating that our results accurately describe the high cyclotron frequency limit relevant to astrophysical and heliospheric systems.   Panels $a$-$c$ show the evolution of $\delta B_j^2/B_0^2$ ($\equiv <(B_j-<B_j>)^2>/B_0^2$; solid) and $B_j^2/B_0^2$ ($\equiv <B_j>^2/B_0^2$; dotted), respectively ($j=x, y,$ and $z$ correspond to black, red, and green, respectively). For the same runs, panels $d$-$f$ show $\Delta p/p_{||}$ ($\equiv <(p_{\perp}-p_{||})/p_{||}>$; black-solid), compared with the linear mirror threshold for a pair plasma and growth rates $\gamma=0.05 \omega_{c,i}$ (panel $d$) and $\gamma=0.007 \omega_{c,i}$ (panels $e$ and $f$). Panels $g$-$i$ show $\delta B_{||}^2/B^2$ and
  $\delta B_{\perp}^2/B^2$. 
  In contrast with the $\beta_{init}=6$ case (shown in Figure \ref{fig:beta6}), here $\delta B_{||}^2 \gg \delta B_{\perp}^2$, which is consistent with the dominance of the mirror instability. Also note that the nonlinear saturation of the mirror fluctuations occurs when $|\delta \vec{B}| \sim 0.3 B$.}
\label{fig:beta20B2s}
\end{figure*}

\noindent The breaking of local $\mu$-invariance for this calculation ($\beta_{\perp}$, $\beta_{||} < \beta_{init}=6$) might be affected by the subdominant (but still significant) contribution of the IC modes to the magnetic fluctuations. However, Figure \ref{fig:muderivs} shows that the magnetic moment varies at the same rate as the mean magnetic field for all of the $\beta_{init}$ we have simulated.   We will discuss the higher $\beta_{init}$ results in more detail in the next two subsections. \newline 
\begin{figure}  \centering \includegraphics[width=9cm]{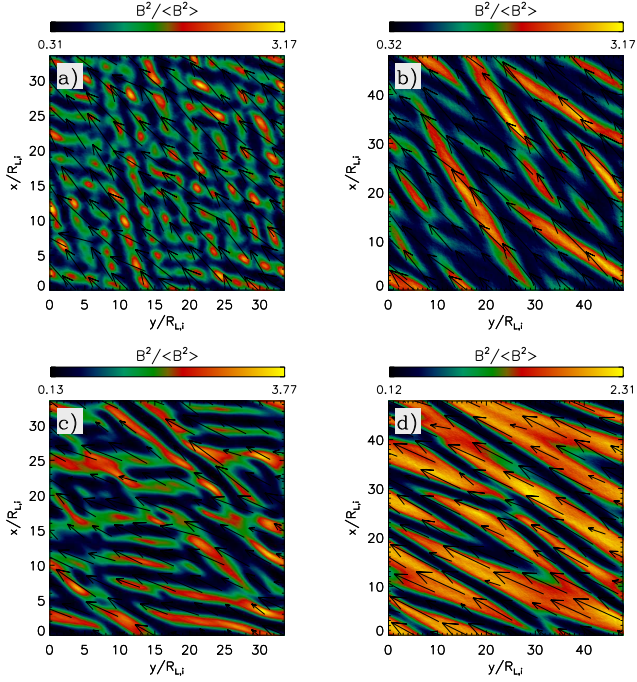}
  \caption{Spatial distribution of magnetic field fluctuations $B^2/<B^2>$ at $t\cdot s \approx 1$ for $\beta_{init} = 20$ and magnetization $\omega_{c,i}/s = 93$ (panel $a$; run beta20mag93b) and $\omega_{c,i}/s = 670$ (panel $b$; run beta20mag670b).  Panels $c$ and $d$ show the same quantities at $t\cdot s \approx 2$.  The black arrows show the direction of $<\vec{B}>$. The higher magnetization run produces mirror modes whose wavevectors are more perpendicular to $<\vec{B}>$ and with longer wavelengths.  A migration to larger wavelengths in time can also be seen.}
\label{fig:fieldsbeta20}
\end{figure}

\noindent An interesting question is whether the non-linear evolution of the velocity space instabilities makes the energy distribution of the particles substantially different from a bi-Maxwellian spectrum. This question is particularly relevant in terms of understanding the suppression of the IC modes in the nonlinear regime as $\beta_{\perp}$ goes from 6 to $\sim 1$. Indeed, it has been shown\citep{Isenberg12,Isenberg13} that a significant departure from a bi-Maxwellian distribution can increase the anisotropy threshold for the growth of the IC modes and that a change of this kind is expected given the resonant character of the IC instability.  Figure \ref{fig:spectrabeta6} shows that the particle spectrum does differ somewhat from a bi-Maxwellian for our $\beta_{init} = 6$ run, mainly due to a significant excess of particles at the highest energies. In future work, we will study whether or not the spectral variation in Figure \ref{fig:spectrabeta6} can completely account for the decrease in the amplitude of the IC modes in the saturated state.  Figure \ref{fig:spectrabeta6} also shows that the ion velocity distribution does not deviate significantly from a bi-Maxwellian at low $v_\parallel$, as has been found in some 1D studies of the saturation of the mirror instability (e.g., \citealt{Southwood93, CalifanoEtAl08}).  We discuss this in more detail in the Appendix.\newline

\noindent Finally, Figure \ref{fig:anistress} shows the evolution of the total heating rate of the particles as a function of time (black line). For comparison, in the red line we plot the particle heating rate by the anisotropic stress, $-s\Delta p\hat{b}_x\hat{b}_y$  \citep{SharmaEtAl06}. We see a reasonable agreement between the two results, implying that the anisotropic stress plays the dominant role in particle heating. Other mechanisms, like wave-particle interaction, must play a secondary role in the total particle energization.  This is true for all of the $\beta_{init}$ we have simulated.   

\subsection{Case $\beta_{init}=20$}

We now consider a somewhat weaker magnetic field case with $\beta_{init}=20$. In this case, the field fluctuations are dominated by the mirror instability at all times. This can be seen in Figures \ref{fig:beta20B2s}$a$-$c$ which show the evolution of the three components of $\delta \vec{B}^2$ for $\beta_{init} = 20$ runs with magnetizations $\omega_{c,i}/s$=93, 670, and 2000, respectively (runs beta20mag93$a$, beta20mag670$a$ and beta20mag2000$a$ in Table 1). In all three cases $\delta B_z^2$ is small compared to $\delta \vec{B}^2$, showing the subdominant role of the IC instability. Figures \ref{fig:beta20B2s}$a$-$c$ also show the characteristic transition between exponential growth (with $\gamma \approx 10 s$) and secular growth (with $\delta \vec{B}^2/B^2 \sim 0.1$) for the mirror modes.  \newline

As in the $\beta_{init}=6$ case, the transition between the exponential and secular regimes occurs at smaller amplitudes and in a smoother way in the more strongly magnetized runs.  Figures \ref{fig:beta20B2s}$d$-$f$ also show that, as in the case of $\beta_{init}=6$, after the overshoot the pressure anisotropy follows the marginal stability condition for the mirror modes reasonably well, particularly at higher magnetization.\newline 

\noindent Figures \ref{fig:beta20B2s}$g$-$i$ show $\delta \vec{B}_{||}^2/B^2$ and $\delta \vec{B}_{\perp}^2/B^2$ as a function of time. While $\delta \vec{B}_{||}^2/B^2$ is about the same for the three magnetizations, $\delta \vec{B}_{\perp}^2/B^2$ is significantly smaller in the higher magnetization runs. The relative magnitude of $\delta \vec{B}_{||}$ and $\delta \vec{B}_{\perp}$ is a measure of the orientation of the mirror's dominant wave vector, $\vec{k}$, with respect to $<\vec{B}>$:  a large $\delta \vec{B}_{||}^2/\delta \vec{B}_{\perp}^2$ ratio implies that  $\vec{k}$ and $<\vec{B}>$ are nearly perpendicular. The results of Figures \ref{fig:beta20B2s}$g$-$i$ imply that for larger magnetization, $\vec{k}$ and $<\vec{B}>$ are more perpendicular, which is consistent with linear calculations \citep{PokhotelovEtAl04}.\newline

\begin{figure}
  \centering \includegraphics[width=9cm]{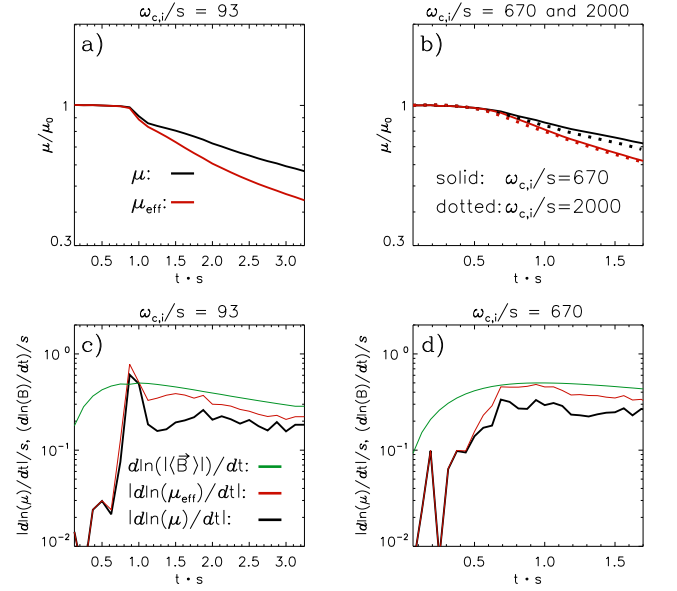} \caption{Panels $a$ and $b$ show the evolution of the ion magnetic moment $\mu$ for $\beta_{init} = 20$ runs with $\omega_{c,i}/s = 93$ (beta20mag93a; panel $a$) and $\omega_{c,i}/s = 670$ and $2000$ (beta20mag670a and beta20mag2000; panel $b$).  Panels $c$ and $d$ compare the corresponding rate of change of $\mu$ with that of the mean magnetic field for runs beta20mag93a and beta20mag670a, respectively.  All plots show the evolution of the true average magnetic moment $\mu$ $\equiv <v_{\perp}^2/B>_p$ (black line) and an effective global magnetic moment, $\mu_{\rm eff}$ $\equiv <v_{\perp}^2>_p/|<\vec{B}>|$ (red line), where the subscript $p$ denotes an average over all particles.  The magnetic moment begins to decrease when the fluctuations reach $\delta B \sim 0.1 \langle B \rangle$ (see Fig. \ref{fig:beta20B2s}).  The difference between $\mu$ and $\mu_{\rm eff}$ is more significant than for $\beta_{init} = 6$ (cf Fig. \ref{fig:beta6b}).  This is a consequence of particles bunching in mirrors, which leads to a correlation between $v_\perp$ and $B$.  Nonetheless, the ion magnetic moment changes at about the same rate as the background magnetic field after saturation at $t s \simeq 0.6$, indicating that pitch angle scattering regulates the nonlinear saturation on sufficiently long timescales.}  \label{fig:mubeta20} \end{figure}

\noindent The orientation of the modes in the nonlinear stage can also be seen directly in Figure \ref{fig:fieldsbeta20}, which shows the spatial distribution of $\vec{B}^2/\langle B^2 \rangle$ at two different times for the runs with $\omega_{c,i}/s=93$ and $\omega_{c,i}/s=670$. Figures \ref{fig:fieldsbeta20}$a$ and \ref{fig:fieldsbeta20}$b$ correspond to the time $t\cdot s=1$, while Figures \ref{fig:fieldsbeta20}$c$ and \ref{fig:fieldsbeta20}$d$ correspond to $t\cdot s=2$. For each magnetization, the mirror fluctuations are dominated by two modes that are symmetric with respect to the magnetic field, and that are more oblique at higher magnetization.   Note also that the wavelength of the dominant modes (in units of the Larmor radius of the particles, $R_{L,i}$) is larger for larger magnetization and that the modes tend to grow in wavelength as time goes on.\newline

\noindent  Figure \ref{fig:mubeta20} shows the evolution of the ion magnetic moment $\mu$ and the rate of change of $\mu$ relative to that of the mean magnetic field for our $\beta_{init} = 20$ calculations.  
A comparison of Figures \ref{fig:mubeta20} \& \ref{fig:beta20B2s} shows that the magnetic moment is reasonably conserved until the magnetic fluctuation amplitude is close to (though somewhat less than) its saturated value. As in our $\beta_{init} = 6$ calculations, this implies that there is a time lag of $\sim 0.25-0.3 s^{-1}$ between the onset of the mirror instability and the onset of significant pitch-angle scattering (and its associated decrease in $\mu$).  Note, however, that for $\beta_{init}=20$ the saturated amplitude of the fluctuations in the secular regime, $|\delta \vec{B}| \sim 0.3 B$, is larger than in the $\beta_{init}=6$ case, for which $|\delta \vec{B}| \sim 0.1 B$.  This implies that when {\it only} the mirror instability plays a significant role, a somewhat larger fluctuation amplitude is necessary for efficient pitch-angle scattering. We will see below that $|\delta \vec{B}| \sim 0.3 B$ continues even for $\beta_{init}=80$;  the saturated amplitude of the mirror modes is thus fairly independent of the beta of the plasma as long as the IC instability is not significant ($\beta \gtrsim$ 6).  \newline

\noindent Figure \ref{fig:mubeta20} also shows that for $\beta_{init} = 20$, the true ion magnetic moment $\mu$ decreases somewhat more slowly than the effective global magnetic moment $\mu_{\rm eff}$ (by $\sim 25 \%$).  This was not seen in our $\beta_{init} = 6$ calculations (Figs. \ref{fig:beta6b}).  The modest difference between $\mu$ and $\mu_{\rm eff}$ indicates that the nonlinear saturation of the mirror instability involves correlations between $v_\perp$ and $B$, i.e., particles bunching in mirrors. This effect is independent of magnetization, as can be seen from the solid and dotted lines in Figure \ref{fig:mubeta20}$b$ (corresponding to $\omega_{c,i}/s=$ 670 and 2000, respectively).\newline

\begin{figure}
\centering
\includegraphics[width=9cm]{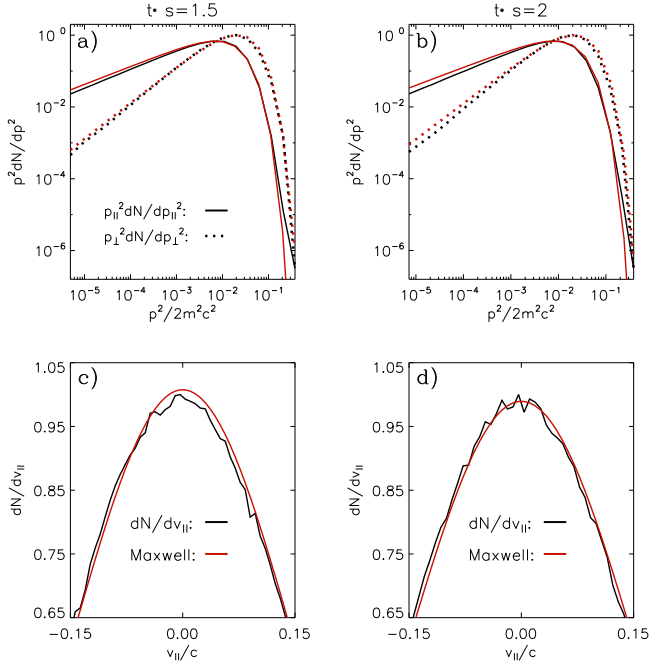}
\caption{Energy spectra of particles for our high magnetization $\beta_{init} = 20$ run beta20mag670a at two different times during the nonlinear regime.  {\em Top:}  Energy spectra as a function of the energies perpendicular and parallel to the magnetic field, $p_{\perp}^2/2m$ (black-dotted line) and $p_{||}^2/2m$ (black-solid line).  {\em Bottom:}   Energy spectra as a function of $v_\parallel$.  In all cases, a bi-Maxwellian energy spectrum (red) provides a good approximation to the numerical simulations.}
\label{fig:spectrabeta20}
\end{figure}

\noindent Figure \ref{fig:spectrabeta20} shows that the particle energy spectrum stays fairly close to bi-Maxwellian in our $\beta_{init} = 20$ runs, regardless of the magnetization. Since these simulations are almost purely mirror dominated, this suggests that the mirror modes provide rather momentum-independent pitch-angle scattering to the particles. As a result, it is likely that IC modes produce the deviations from a bi-Maxwellian distribution seen in Figure \ref{fig:spectrabeta6}.  Figure \ref{fig:spectrabeta20} also shows that the parallel velocity spectrum $dN/dv_\parallel$ remains nearly Maxwellian during most of the saturated regime (with only a transient $\sim 1\%$ decrease for $v_{||} \to 0$ especially at the end of the exponential growth regime). The lack of significant flattening of $dN/dv_\parallel$ \citep[which contrasts with the prediction of][]{CalifanoEtAl08} is analyzed in further detail in the Appendix.

\subsection{Case $\beta_{init}=80$}

\begin{figure}
\centering
\includegraphics[width=9cm]{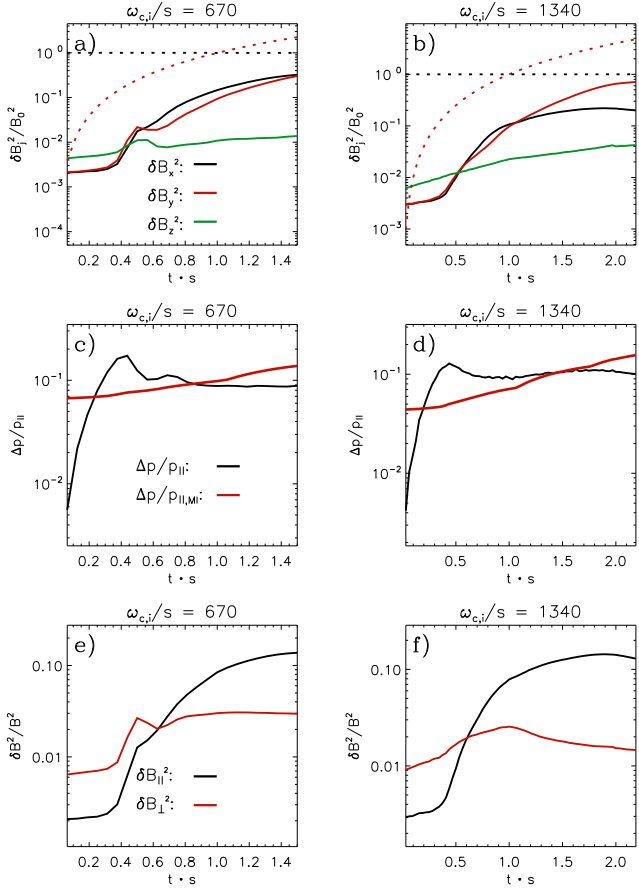}
\caption{Magnetic field fluctuations and pressure anisotropy for $\beta_{init} = 80$ runs with $\omega_{c,i}/s = 670$ (beta80mag670; left column) and $\omega_{c,i}/s = 1340$ (beta80mag1340; right column). Panels $a$ and $b$ show the evolution of $\delta B_j^2/B_0^2$ ($\equiv <(B_j-<B_j>)^2>/B_0^2$; solid) and $B_j^2/B_0^2$ ($\equiv <B_j>^2/B_0^2$; dotted), respectively ($j=x, y,$ and $z$ correspond to black, red, and green, respectively). For the same runs, panels $c$ and $d$ show $\Delta p/p_{||}$ ($\equiv <(p_{\perp}-p_{||})/p_{||}>$; black-solid), compared with the linear mirror threshold  in red ($\Delta p/p_{||,MI}$) for a pair plasma and growth rates $\gamma=0.007 \omega_{c,i}$ and $\gamma=0.0035 \omega_{c,i}$, respectively. Panels $e$ and $f$ show $\delta B_{||}^2/B^2$ and
$\delta B_{\perp}^2/B^2$ .
The results shown here are very similar to the analogous $\beta_{init}=20$ results in Figure \ref{fig:beta20B2s}, consistent with the dominance of the mirror instability at high $\beta$.}
\label{fig:beta80B2s}
\end{figure}

\noindent For $\beta_{init}=80$ we focus on simulations with $\omega_{c,i}/s=670$ and 1340, respectively (runs beta80mag670 and beta80mag1340 in Table 1). The results are essentially the same as for the $\beta_{init}=20$ case so we do not discuss them in detail. Figure \ref{fig:beta80B2s} shows that the mirror instability dominates the fluctuations in $\vec{B}$. The evolution of the pressure anisotropy $\Delta p/p_{||}$ contains the initial overshoot followed by a rather flat behavior. The amplitude of the overshoot is controlled by the magnetization, while the subsequent evolution is fairly well described by the threshold condition for mirror modes.\newline 

\begin{figure}
  \centering \includegraphics[width=9cm]{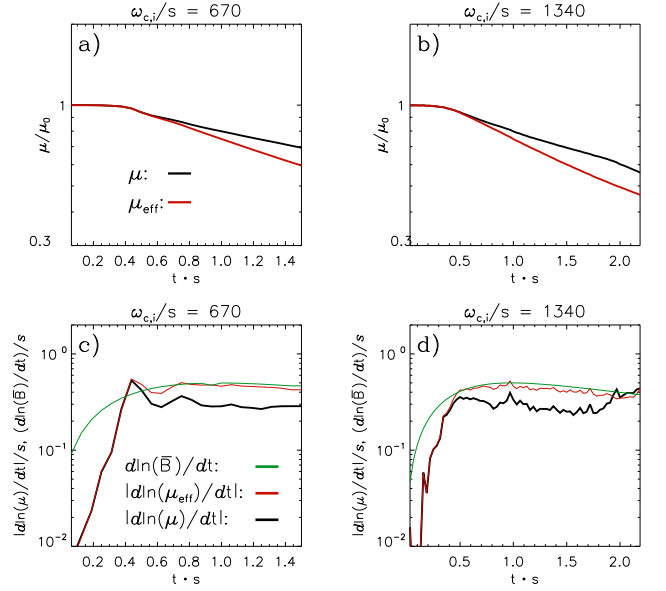} \caption{Panels $a$ and $b$ show the evolution of the ion magnetic moment $\mu$ for our $\beta_{init} = 80$ runs with $\omega_{c,i}/s = 670$ and $\omega_{c,i}/s = 1340$ (runs beta80mag670 and beta80mag1340, respectively).  Panels $c$ and $d$ compare the rate of change of $\mu$ with that of the mean magnetic field.  In all panels, we show the evolution of the true average magnetic moment $\mu$ $\equiv <v_{\perp}^2/B>_p$ (black line) and an effective global magnetic moment, $\mu_{\rm eff}$ $\equiv <v_{\perp}^2>_p/|<\vec{B}>|$ (red line), where the subscript $p$ denotes an average over all particles.  The magnetic moment begins to decrease when the fluctuations reach $\delta B \sim 0.1 \langle B \rangle$ (see Fig. \ref{fig:beta80B2s}).  All of the key results here are very similar to the $\beta_{init}=20$ results shown in Figure \ref{fig:mubeta20}.}
\label{fig:mubeta80}
\end{figure}

\noindent As in the $\beta_{init}=20$ case, the saturation $\delta B/B$ is $\sim 0.3$. Figure \ref{fig:mubeta80} shows the evolution of the ion magnetic moment for the $\beta_{init} = 80$ simulations.  The evolutions of $\mu$ and $\mu_{eff}$ are the same for the two magnetizations presented ($\omega_{c,i}/s = 670$ and $\omega_{c,i}/s = 1340$), and show the same behavior seen in the $\beta_{init}=20$ case, which is discussed in detail in the previous subsection.\newline 

\subsection{Comparison with $m_i/m_e>1$ simulations} \label{sec:compmime} \begin{figure}
  \centering \includegraphics[width=9cm]{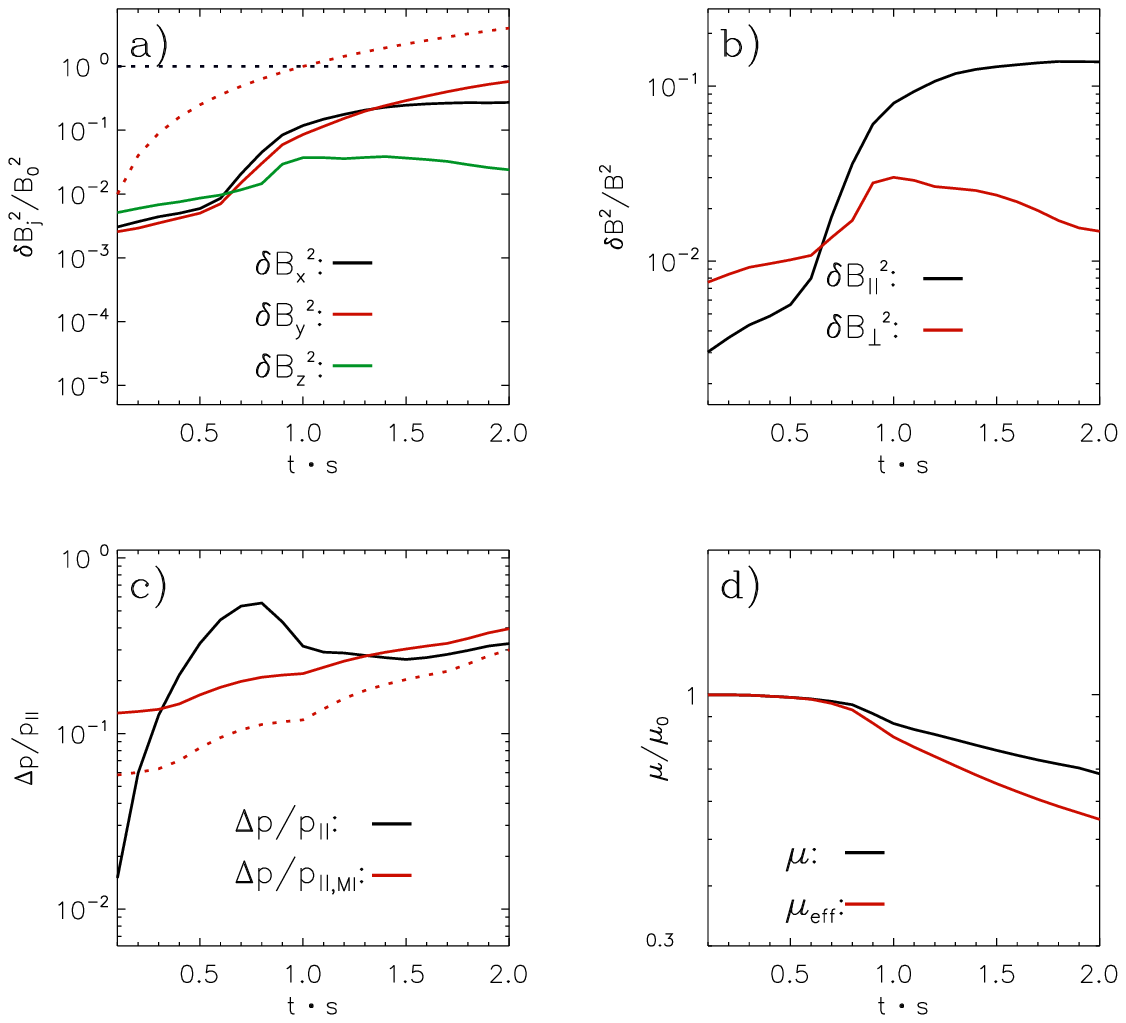} \caption{Magnetic field fluctuations, ion pressure anisotropy, and ion magnetic moment for a simulation with $m_i/m_e=10$, $\beta_i=\beta_e=20$, and $\omega_{c,i}/s=240$ (run betai20magi240mass10).  The ion results here are similar to the corresponding $m_i = m_e$ results in Figures \ref{fig:beta20B2s} and \ref{fig:mubeta20}, indicating that the pair plasma results capture most of the key ion physics.  Panel $a$ shows the evolution of $\delta B_i^2/B_0^2$ and $B_i^2/B_0^2$, in solid and dotted lines, respectively. 
The small contribution of $\delta B_z^2$ implies the dominance of mirror modes. Panel $b$ shows the evolution of $\delta B_{\perp}^2/B^2$ (red) and $\delta B_{||}^2/B^2$ (black) with $\delta B_{||}^2 \gg \delta B_{\perp}^2$, implying that the mirror modes have $k_{\perp} \gg k_{||}$. Panel $c$ shows the evolution of the ion pressure anisotropy $\Delta p_i/p_{||,i}$ (black-solid). The linear mirror threshold for mirror modes with $\gamma/\omega_{c,i}=0.05$ in a $\beta_i=\beta_e$, $m_i/m_e=10$ plasma is also shown (red line) along with the analogous threshold for $m_i/m_e = 1836$ for comparison (red dotted line).  Panel $d$ shows the evolution of $\mu$ ($\equiv <v_{\perp,i}^2/B>$, black-solid) and $\mu_{\rm eff}$ ($\equiv <v_{\perp,i}^2>/<B>$, red-solid); see eq. \ref{eq:mu} and associated text for details.}
\label{fig:beta20B2smime4}
\end{figure}

\noindent In this section we show that our use of $m_i/m_e=1$ in the previous sections does not have a significant effect on the evolution of the ion pressure anisotropy under the influence of the IC and mirror instabilities. In Figure \ref{fig:beta20B2smime4} we present the results for a simulation with $m_i/m_e=10$, $\beta_i=\beta_e=20$, and $\omega_{c,i}/s=240$ (run betai20magi240mass10c), which is analogous to the previous $m_i/m_e=1$ simulations that use $\beta_i=\beta_e=10$. A comparison of Figure \ref{fig:beta20B2smime4}$a$ with Figures \ref{fig:beta20B2s}$a$, \ref{fig:beta20B2s}$b$, and \ref{fig:beta20B2s}$c$ shows that for both mass ratios the evolution of $\delta \vec{B}$ is dominated by the mirror instability, with the initial exponential regime being followed by secular growth. Figure \ref{fig:beta20B2smime4}$b$ shows the evolution of $\delta B_{\perp}^2/B^2$ and $\delta B_{||}^2/B^2$. As in the $m_i/m_e=1$ case (see Figures \ref{fig:beta20B2s}$g$, \ref{fig:beta20B2s}$h$, and \ref{fig:beta20B2s}$i$), $\delta B_{||}^2 \gg \delta B_{\perp}^2$ in the secular stage, implying that mirror modes with $k_{\perp} \gg k_{||}$ are dominant. The maximum amplitude of $\delta B_{||}/B \sim 0.3$ is also in good agreement with the $m_i/m_e=1$ simulations.\newline 

\noindent The evolution of the ion pressure anisotropy $\Delta p_i/p_{||,i}$ is shown in Figure \ref{fig:beta20B2smime4}$c$. As in the $m_i/m_e=1$ runs (see Figures \ref{fig:beta20B2s}$d-f$), $\Delta p_{i}/p_{||,i}$ follows the linear threshold of the mirror instability fairly closely. This is seen in Figure \ref{fig:beta20B2smime4}$c$, which shows the linear mirror threshold expected for modes with $\gamma/\omega_{c,i}=0.05$, $\beta_i=\beta_e=20$, and $m_i/m_e=10$ and 1836 (red and dotted-red lines, respectively). The small difference between our results and the $m_i/m_e=10$ and 1836 thresholds strongly suggests that the use of $m_i/m_e=1$ for most of our study captures reasonably well the key physics of the mirror and IC instabilities in a real electron-ion plasma. Finally, Figure \ref{fig:beta20B2smime4}$d$ shows the evolution of $\mu$ (black solid line) and $\mu_{\rm eff}$ (red solid line).  The results are quantitatively similar to the $m_i = m_e$ results in Figures \ref{fig:mubeta20}$a$ and \ref{fig:mubeta20}$b$. This shows that the role played by pitch-angle scattering in the saturation of the mirror instability is well captured by our $m_i/m_e=1$ simulations.

\section{Conclusions}
\label{sec:conclu}

\noindent We have used PIC simulations to study the nonlinear, long-term evolution of ion velocity space instabilities in collisionless plasmas. We have focused on instabilities driven by pressure anisotropy with $p_{\perp,i} > p_{\parallel,i}$:  the ion cyclotron (IC) and mirror instabilities. These instabilities are expected to arise when turbulence and/or MHD instabilities amplify $\vec{B}$, rendering $p_{\perp,i} > p_{\parallel,i}$ due to conservation of the ion magnetic moment $\mu_i$.  Alternatively, in the solar wind $p_{\perp,i}/p_{\parallel,i}$ can increase due to perpendicular heating and/or parallel cooling of the plasma.\newline

\noindent In contrast to previous studies, we do not consider the initial value problem of the evolution of a given initial ion pressure anisotropy. Instead, we self-consistently induce the growth of an ion pressure anisotropy by continuously amplifying the mean magnetic field in our computational domain. This setup allows us to study the long-term, saturated state of the plasma, rather than the (much shorter) initial, exponential stage.  The former is of most interest for the heliospheric and astrophysical applications of our work. We have focused on the regime $\beta_i \approx 1-80$ and plasma magnetizations $\omega_{c,i}/s$ in the range $\omega_{c,i}/s \sim 100-1000$
(the magnetization is the ratio of the initial ion cyclotron frequency to the background shear frequency).  Most of our simulations used $m_i/m_e=1$, but we showed that our results for the ion scale physics are fairly independent of the mass ratio (see \S \ref{sec:compmime}).  So long as $\omega_{c,i}/s \gg 1$, our results are also relatively independent of the precise value of the magnetization (see, in particular, Figs. \ref{fig:beta20B2s} \& \ref{fig:mubeta20}). Thus we believe that our basic conclusions about the saturation of the mirror and IC instabilities in continuously driven systems can be applied to physical problems where $\omega_{c,i}$ is many orders of magnitude larger than $s$.\newline

\noindent Our primary results are as follows:
\begin{itemize}
\item For the $\beta_i \approx 1-80$ regime we have studied, the turbulent fluctuations in the saturated state are dominated by mirror modes. This is despite the fact that IC and mirror instabilities are both present at early phases (and have similar linear instability thresholds) when $\beta_i \approx$ {a few}.  A plausible explanation for this dominance of the mirror instability in the saturated state even at $\beta_i \sim$ a few is that the ion velocity distribution departs from a bi-Maxwellian spectrum in the presence of the IC instability.    This can in turn increase the threshold for the IC instability \citep{Isenberg12,Isenberg13} making it less important than the mirror instability in the nonlinear regime. \footnote{The mirror instability can dominate the dynamics even more when the anisotropic pressure of the electrons \citep{RemyaEtAl13} and the full 3D geometry of the problem \citep{ShojiEtAl09} are taken into consideration.}  We indeed see more significant departures from a bi-Maxwellian in the low $\beta_i$ regime where the IC instability is present (Figs. \ref{fig:spectrabeta6} \& \ref{fig:spectrabeta20}).   By contrast, the ion distribution function remains relatively bi-Maxwellian in the high $\beta_i$ regime in which the mirror instability dominates over the IC instability at all times (Fig. \ref{fig:spectrabeta20}). The dominance of the mirror instability at $\beta_i \approx$ a few is consistent with the nonlinear study of \cite{TravnicekEtAl07}, which makes use of hybrid simulations of an expanding box.

\item Small-scale fluctuations driven by the mirror instability initially grow exponentially but saturate in a secular phase in which $|\delta \vec{B}| \sim 0.2-0.3 |<\vec{B}>|$ (Figures \ref{fig:beta6} \& \ref{fig:beta20B2s}).  The fluctuations in the saturated state have $k_{\perp} \gg k_{||}$ (see, e.g., Figure \ref{fig:fieldsbeta6}) consistent with linear theory expectations for mirror modes.

\item The ion pressure anisotropy $\Delta p_i/p_{||,i}$ in the saturated state is reasonably well described by the marginal stability condition for the mirror instability (e.g., Figs \ref{fig:beta6} \& \ref{fig:beta20B2s}).   Note, however, that the marginal stability state of the mirror mode in our simulations is not the same as in real systems because of the comparable ion and electron masses and the smaller $\omega_{c,i}/s$ used in our calculations. This is important to bear in mind when comparing our results to heliospheric measurements.

\item The total thermal energy of the plasma increases in time at the theoretically predicted rate given the background shear in the plasma and the anisotropic stress associated with the pressure anisotropy (Fig. \ref{fig:anistress}).  This is consistent with the idea that the ``viscous'' heating rate in a collisionless plasma is set by how velocity space instabilities regulate the pressure anisotropy \citep{SharmaEtAl06}.

\item When the growth time of the mirror instability is long compared to the ion cyclotron period -- as is the case in our simulations where the shear slowly amplifies the mean magnetic field -- the ion magnetic moment is initially reasonably well conserved. This is fundamentally because there is no mechanism to violate magnetic moment conservation during the linear phase of mirror growth. So the modes continue to grow to larger and larger amplitudes, given the sustained free energy source created by the background velocity shear amplifying the magnetic field and continuously generating pressure anisotropy. This regime lasts until $\delta B \sim 0.1 \, \langle B \rangle$ and $ts \lesssim 1$, and is consistent with the ``secular regime"  found by \citet{KunzEtAl14} where $\mu$ is nearly conserved.

\item Saturation of the mirror modes starts to happen once the fluctuations attain $\delta B \sim 0.1 \, \langle B \rangle$. In this stage the variation in magnetic field on the scale of the larmor radius is sufficiently large that the ion magnetic moment is no longer conserved. This is consistent with theories of stochastic ion motion in large amplitude turbulent fluctuations \citep[e.g.,][]{ChenEtAl01, JohnsonEtAl01}. In this phase the ion magnetic moment changes on nearly the same timescale as the mean magnetic field (Fig. \ref{fig:muderivs}), which is consistent with pitch angle scattering maintaining marginal stability. In this stage the magnetic fluctuations only experience a modest additional growth, reaching a maximum amplitude of $\delta B \sim 0.3 \, \langle B \rangle$.

\item The need for finite amplitude fluctuations with $\delta B \sim 0.1 \, \langle B \rangle$ in order to stop magnetic moment conservation implies the existence of a the temporal delay of $\sim 0.3 s^{-1}$ between the onset of the IC and mirror instabilities and the onset of efficient pitch angle scattering ($s$ is the rate at which the mean magnetic field is growing).  
\end{itemize}

\noindent Our results have a number of consequences for modeling the large scale dynamics of nearly collisionless astrophysical plasmas and for interpreting heliospheric measurements.  We briefly mention a few of these applications but defer a more detailed analysis to future work.\newline

\noindent Our results are consistent with a number of in situ heliospheric measurements. For instance, our maximum saturated value of $\delta B/B \sim 0.3$ is in reasonable agreement with mirror modes observations in different environments like planetary magnetosheaths \citep[where large amplitude mirror modes are frequently found. See, e.g.,][]{TsurutaniEtAl82, JoyEtAl06, SoucekEtAl08, VolwekEtAl08, HorburyEtAl09,TsurutaniEtAl11a}, the solar wind \citep[e.g.,][]{LiuEtAl06, bale2009, RussellEtAl09, Enriquez-riveraEtAl10}, and the heliosheath \citep[e.g.,][]{TsurutaniEtAl11a,TsurutaniEtAl11b}.\footnote{We note that our quoted maximum $\delta B/B\sim 0.3$ corresponds to $<(\vec{B}-< \vec{B}>)^2>^{1/2} / |<\vec{B}>|$, where $< >$ stands for volume average. This means that our results still allow for {\it local} magnetic fluctuations of even larger amplitude. It is also important to stress that if a plasma is rapidly driven to have a temperature anisotropy well in excess of the mirror threshold \citep[e.g., at the Earth's bow shock;][]{TsurutaniEtAl82} then the resulting amplitude of the mirror modes could significantly exceed what we find in our calculations.    In this case, a standard initial value calculation of the mirror evolution is likely to better capture the resulting dynamics than our model in which the anisotropy slowly increases in time.} Our results also show that the dominant wavelength is initially close to a few ion Larmor radii, with a subsequent migration to longer wavelengths through a coalescence process. This wavelength growth is consistent with observations \citep[e.g.,][]{TsurutaniEtAl11b, SchmidEtAl14} and with theoretical models that predict the growth of mirror mode wavelengths due to magnetic diffusion \citep{HasegawaEtAl11}.\newline

\noindent The fact that our $\beta_i \sim 1-6$ calculations find that the mirror instability dominates the nonlinear state of the fluctuations (even though the IC and mirror instabilities have comparable linear thresholds and growth rates for a bi-Maxwellian distribution) is also consistent with heliospheric observations. In particular, solar wind measurements at $\beta_i\sim 1-10$ show that $p_{\perp,i}/p_{||,i}$ is limited by the linear threshold of the mirror instability \citep[instead of the IC instability. See ][]{hellinger2006,bale2009}. The observations of \citet{bale2009} also show enhanced magnetic fluctuations and magnetic compressibility that follow the linear mirror threshold fairly well. It is important to emphasize, however, that the observations presented by \citet{hellinger2006} and \citet{bale2009} showing dominance of the mirror instability also include the regime $\beta_i \ll 1$, which is not considered in our study. Also, the dominance of the mirror instability could be affected by properties of the solar wind that are usually not included in calculations of instability thresholds. These include the presence of $\sim 5\%$ of alpha particles \citep[see, e.g., Figure 7 of ][]{MatteiniEtAl12}, the drift of alpha particles with respect to protons \citep{MarschEtAl82a}, and the frequent presence of a proton beam \citep{MarschEtAl82b}.\newline 

\noindent We interpret the dominance of the mirror modes in our simulations as being due to the suppression of the IC waves, which is caused by the ion velocity distribution function departing from bi-Maxwellian \citep{Isenberg12,Isenberg13}. This interpretation also requires the IC modes to have finite amplitudes (although smaller than the one of the mirror modes) so that they can maintain, via pitch-angle scattering, an ion distribution function that is marginally stable to the growth of IC waves.  To the best of our knowledge, these IC modes have not been found by in-situ heliospheric measurements during mirror dominated events. This is consistent with the fact that in our simulations the mirror and IC modes only have comparable amplitudes during the initial, exponential regime (Fig. \ref{fig:beta6}$a$ and $b$). In the saturated regime (which is likely to be more representative of the observed state of the plasma in the heliospheric and astrophysical systems), we find that the mirror modes have significantly larger amplitudes. Moreover, this dominance increases with the magnetization  $\omega_{c,i}/s$, as can be seen by comparing Figures \ref{fig:beta6}$a$ and $b$. \footnote{This dependence on $\omega_{c,i}/s$ can be accounted for by the fact that the scattering rate of cyclotron-frequency fluctuations is expected to be $\sim \omega_{c,i} (\delta B/B)^2$. If we equate this to the shear parameter $s$ (which is, roughly, the rate at which pitch-angle scattering takes place, as can be seen from the changing rate of $\mu$ in Fig. \ref{fig:muderivs}), one obtains that $(\delta B/B)^2 \sim s/\omega_{c,i}$.} Thus, for realistic values of $\omega_{c,i}/s$, we do not anticipate easily detectable IC waves by in-situ measurements in cases where the mirror modes dominate. The departure from the bi-Maxwellian ion distribution function has not been confirmed by in-situ heliospheric measurements of mirror-dominated events either. We believe that this is due to the high sensitivity of resonant instabilities to gradients in the velocity distribution function. Thus, in cases where IC and mirror modes can have similar growth rates, only small deviations from the bi-Maxwellian distribution can significantly reduce the importance of the IC modes and give way to the dominance of the mirror modes. We, therefore, do not anticipate easily detectable departures from bi-Maxwellian associated to this effect. \footnote{We notice that Helios measurements show distribution functions with significant deviations from the bi-Maxwellian shape in the fast solar wind, consistent with predictions based on the presence of IC waves \citep{TuEtAl02, HeuerEtAl07}. Although these measurements do not correspond to the small deviations needed to suppress the IC modes that we consider here, they do show the strong effect that high amplitude IC modes can have on the ion distribution.} Nevertheless, future careful observations comparing the nature of the fluctuations and the ion distribution functions in the solar wind can help to unveil these effects.\newline

\noindent A second application of our results is to low-collisionality accretion flows onto black holes. In these systems pressure anisotropies can produce an anisotropic stress that can increase the transport of angular momentum and the heating of the plasma \citep{SharmaEtAl06}. Our results are consistent with \citet{SharmaEtAl06}'s hypothesis that the mirror and/or IC instabilities would provide enhanced pitch angle scattering and regulate the evolution of the plasma pressure tensor.  Axisymmetric kinetic simulations have also demonstrated this directly \citep{RiquelmeEtAl12} though the restriction to axisymmetry limits the quantitative applicability of these initial simulations to real disks.  One of the key conclusions of our study in this paper is that even modest plasma magnetization is sufficient to quantitatively capture the nonlinear saturation of the mirror instability.  For example, our $\beta_i = 20$ results in Figures \ref{fig:beta20B2s} \& \ref{fig:mubeta20} show that the evolution of the ion pressure anisotropy, ion magnetic moment, and turbulent fluctuations are quite similar for $\omega_{c,i}/s = 93$ and $\omega_{c,i}/s = 670$.  The primary difference is that in the lower magnetization run the plasma has a larger ``overshoot'' of the saturation pressure anisotropy set by the linear mirror instability threshold.  This weak dependence on magnetization is encouraging because it is computationally infeasible to reach $\omega_{c,i}/s = 670$ in three-dimensional kinetic simulations of astrophysical plasmas (e.g., accretion disks).\newline

\noindent One of the interesting questions not fully addressed by our work is the saturation of velocity space instabilities in turbulent low-collisionality plasmas where the shear rate of the mean magnetic field itself fluctuates in time and the magnitude of the large-scale turbulent fluctuations satisfies $\delta B \ll \langle B \rangle$.  Our calculations in this paper have focused on the case where an imposed velocity shear amplifies a background magnetic field by order unity or more.  If, on the other hand, the shear acts for less time such that the background field is amplified by significantly less, the plasma may never become mirror/IC unstable (depending on $\beta_i$ and $\delta B/B$) or the background shear may reverse sign before the mirror/IC fluctuations have reached sufficiently large amplitudes for pitch angle scattering to set in.  This regime will be explored in future work. \newline

\noindent Another key question not addressed by our study is isotropization of the electron pressure anisotropy.
In future work we will study how fast electrons interact with the large-scale mirrors generated by the ion pressure anisotropy, and whether there are separate electron-scale fluctuations that produce significant electron pitch-angle scattering (e.g., the whistler instability). \newline

\acknowledgements

We thank Tobi Heinemann, Ben Chandran, Phil Isenberg and, in particular, Matt Kunz for fruitful discussions.  EQ is supported in part by NASA HTP grant NNX11AJ37G, NSF Grant AST-1333682, a Simons Investigator award from the Simons Foundation, the David and Lucile Packard Foundation, and the Thomas Alison Schneider Chair in Physics. MR thanks support from the Chilean Comisi\'on Nacional de Investigaci\'on Cient\'ifica y Tecnol\'ogica (CONICYT; Proyecto Fondecyt Iniciaci\'on N$^{\textrm{o}}$ 11121145).  DV is supported in part by NASA HTP grant NNX11AJ37G and NASA grant NNX12AB27G.


\appendix
\section{Mirror Saturation in 1D and 2D}

\noindent Previous 1D studies of the mirror instability suggest that its growth and saturation are dominated by the ``resonant" scattering of $v_{||} \to 0$ particles \citep[where $v_{||}$ is the particle velocity component parallel to the magnetic field;][]{Southwood93,CalifanoEtAl08}, leading to a flattening of the distribution function near $v_{||} \to 0$.   In contrast, our 2D runs discussed in the main text show that scattering by mirror modes is fairly momentum independent, with no significant flattening of $dN/dv_{||}$.   To compare our results more directly to the previous literature, in this appendix we compare the mirror saturation in 1D and 2D for different values of $\beta$, paying especial attention to the behavior of $v_{||} \to 0$ particles. We show that, compared with the 1D case, 2D runs significantly change the way pitch-angle scattering occurs, modifying the saturation mechanism.\newline 
\begin{figure*}[t]
\centering
\includegraphics[width=8cm]{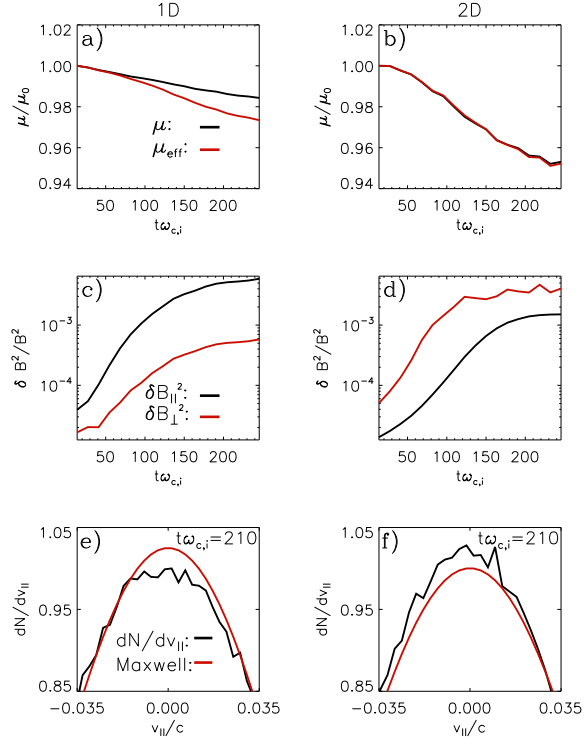} \caption{We compare 1D and 2D initial-value simulations with $m_i=m_e$, $\beta_{\perp,i}=\beta_{\perp,e}=1.15$, $\Delta p_i/p_{||,i}=\Delta p_e/p_{||,e}=1.3$. Apart from the different number of dimensions, the simulations have the same numerical parameters (see below). Panels $a$ and $b$ show $\mu$ and $\mu_{eff}$ for the 1D and 2D simulations, respectively.  Panels $c$ and $d$ show the magnetic energy of the fluctuations that are parallel ($\delta B_{||}^2$) and perpendicular ($\delta B_{\perp}^2$) to the mean magnetic field, normalized in terms of $B^2$. Finally, panels $e$ and $f$ show snapshots of the parallel velocity distribution $dN/dv_{||}$ (black) at $t\omega_{c,i}=210$ in both cases, compared to a bi-Maxwellian (red). The numerical parameter of the simulations are: $[c/\omega_{p,e}]/\Delta_x=10$, $N_{ppc}=140$, and $L/R_{L,i}=70$. In 1D, the angle between $\vec{B}_0$ and the direction of the resolved dimension is $73^o$.}  \label{fig:1d2d2.3} 
\end{figure*}

\noindent To best compare with the previous literature, we use initial-value simulations in this Appendix, where a linearly unstable pressure anisotropy is imposed as an initial condition.  Figure \ref{fig:1d2d2.3} compares simulations with $m_i=m_e$, $\beta_{\perp,i}=\beta_{\perp,e}=1.15$, $\Delta p_i/p_{||,i}=\Delta p_e/p_{||,e}=1.3$ in 1D and 2D.  In the 1D simulation, the initial magnetic field $\vec{B}_0$ forms an angle of $73^o$ with the resolved dimension of the simulation.   Apart from the different number of dimensions, these simulations also have the same numerical parameters.  Panels \ref{fig:1d2d2.3}$a$ and \ref{fig:1d2d2.3}$b$ show $\mu$ and $\mu_{eff}$ as a function of time for the 1D and 2D cases, respectively. In 1D, $\mu$ decreases significantly slower than $\mu_{eff}$, which means that there is a significant correlation between $p_{\perp}$ and $B$ (as expected for the mirror instability). In 2D, on the other hand, the difference between $\mu$ and $\mu_{eff}$ is very small, implying that pitch-angle scattering occurs more efficiently in 2D.\newline

\noindent This difference in the scattering efficiency can be explained considering the modes that grow in the 1D and 2D simulations. Indeed, in the 1D case, mirror modes with wave vectors pointing in only one direction can grow. In the 2D case, on the other hand, both IC and mirror modes appear (with the mirror modes forming both positive and negative angles with respect to $\vec{B}_0$). This can be seen from panels \ref{fig:1d2d2.3}$c$ and \ref{fig:1d2d2.3}$d$, which show the evolution of the magnetic energy contained in fluctuations parallel ($\delta B_{||}^2$) and perpendicular ($\delta B_{\perp}^2$) to $\vec{B}_0$ for the 1D and 2D cases. Whereas the 1D case is dominated by $\delta B_{||}^2$ (characteristic of mirror modes), the 2D case is dominated by $\delta B_{\perp}^2$, showing the dominant presence of the IC modes. Panels \ref{fig:1d2d2.3}$e$ and \ref{fig:1d2d2.3}$f$ show snapshots of the parallel velocity distribution $dN/dv_{||}$ for the 1D and 2D runs at the saturated state $t\omega_{c,i}=210$ (black lines), and compare them with a bi-Maxwellian distribution (red line). In the 1D case, we can see flattening of $dN/dv_{||}$ for $v_{||} \to 0$. This flattening is consistent with the results of \cite{CalifanoEtAl08}, who argue that it is due to the resonant interaction of $v_{||} \to 0$ particles with the mirror modes. The 2D run, however, does not have this flattening, implying that the saturation process in two dimensions is quite different. This is indeed expected from the dominant role of the IC instability.\newline
\begin{figure*}[t]
\centering
\includegraphics[width=8cm]{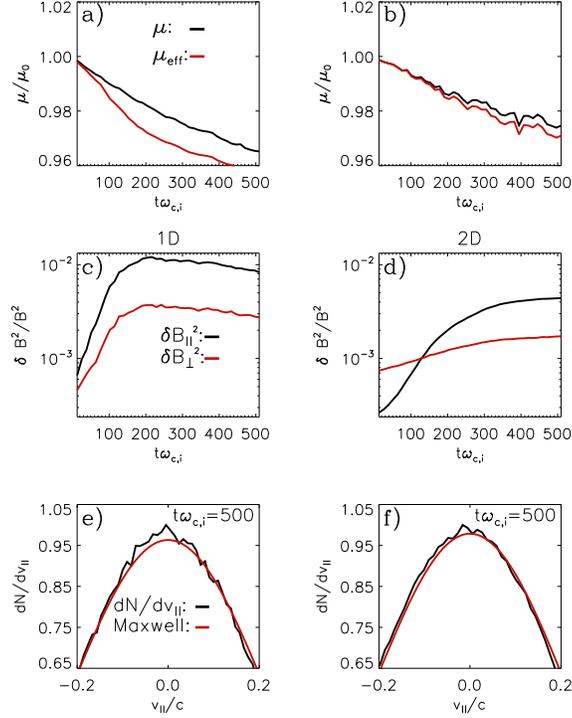} \caption{Same as Figure \ref{fig:1d2d2.3} but for simulations with: $m_i=m_e$, $\beta_{\perp,i}=\beta_{\perp,e}=10$, $\Delta p_i/p_{||,i}=\Delta p_e/p_{||,e}=0.3$, $[c/\omega_{p,e}]/\Delta_x=5$, $N_{ppc}=100$, and $L/R_{L,i}=70$. In 1D, the angle between $\vec{B}_0$ and the direction of the resolved dimension is $62^o$. The snapshot for $dN/dv_{||}$ is taken at $t\omega_{c,i}=500$.}  \label{fig:1d2d20} 
\end{figure*}

\noindent Since at higher $\beta \gtrsim 20$ the IC instability is expected to play a subdominant role, we also explored the case where initially $\beta_{\perp}=20$. Figure \ref{fig:1d2d20} compares simulations with $m_i=m_e$, $\beta_{\perp,i}=\beta_{\perp,e}=10$, and $\Delta p_i/p_{||,i}=\Delta p_e/p_{||,e}=0.3$ in 1D and 2D, with both cases having the same numerical parameters.
Panels \ref{fig:1d2d20}$a$ and \ref{fig:1d2d20}$b$ show the evolution of $\mu$ and $\mu_{eff}$ for both cases. In the 1D case these two quantities differ significantly, implying an important correlation between $B$ and $p_{\perp}$. This correlation almost disappears in the 2D case, which shows the presence of more efficient pitch-angle scattering in the 2D case. Since for $\beta_{\perp}=20$ the IC instability plays a subdominant role, this is likely a property of the mirror instability itself rather than the presence of the IC instability as was the case for the lower $\beta$ results shown in Figure \ref{fig:1d2d2.3}.   
The higher rate of scattering in 2D also affects the maximum amplitude of the fluctuations, which are shown in panels \ref{fig:1d2d20}$c$ and \ref{fig:1d2d20}$d$ for the 1D and 2D cases. Although in both cases $\delta B_{||}^2 > \delta B_{\perp}^2$ (implying that mirror dominates), in 2D the amplitude of the modes is smaller, which means that field rearrangement plays a less important role in the approach to marginal stability.\newline

\noindent Finally, Figure \ref{fig:1d2d20} also compares $dN/dv_{||}$ at the saturated state ($t\omega_{c,i}=500$) in the 1D and 2D cases. There is very little flattening for $v_{||} \to 0$ in both cases. This confirms our result from the main text that the pitch-angle scattering in the mirror dominated case is fairly independent of particle momentum. It is interesting, however, that we do not see flattening even in 1D.  Since \cite{CalifanoEtAl08} reported the appearance of flattening only near the threshold of the mirror instability (and only for certain box sizes), we tested simulations with the same plasma parameters having anisotropies as low as $\Delta p_i/p_{\parallel,i}=\Delta p_e/p_{\parallel,e}=0.15$ and found similar results.  
\end{document}